\documentclass[fleqn,usenatbib]{mnras}

\usepackage[T1]{fontenc}

\DeclareRobustCommand{\VAN}[3]{#2}
\let\VANthebibliography\thebibliography
\def\thebibliography{\DeclareRobustCommand{\VAN}[3]{##3}\VANthebibliography}

\usepackage{graphicx}	% Including figure files
\usepackage{amsmath}	% Advanced maths commands
\usepackage{amssymb}	% Extra maths symbols
\usepackage{xspace}

\newcommand{\water}{H$_2$O\xspace}  % water

\newcommand{\logL}{$\log L$\xspace}
\newcommand{\KpVsys}{$K_\mathrm{p} - V_\mathrm{sys}$\xspace}
\newcommand{\KpVrest}{$K_\mathrm{p} - V_\mathrm{rest}$\xspace}
\newcommand{\Kp}{$K_\mathrm{p}$\xspace}
\newcommand{\Vsys}{$V_\mathrm{sys}$\xspace}
\newcommand{\Vrest}{$V_\mathrm{rest}$\xspace}
\newcommand{\logwater}{$\log_{10}(\mathrm{H_2O})$\xspace}
\newcommand{\logcloud}{$\log_{10}(P_\mathrm{cloud}\mathrm{/bar})$\xspace}

\newcommand{\Mjup}{\ensuremath{\mathrm{M_{J}}}\xspace}
\newcommand{\Rjup}{\ensuremath{\mathrm{R_{J}}}\xspace}
\newcommand{\Mnep}{\ensuremath{\mathrm{M_{Nep}}}\xspace}
\newcommand{\Rnep}{\ensuremath{\mathrm{R_{Nep}}}\xspace}

\newcommand{\ms}{\ensuremath{\mathrm{m\,s^{-1}}}\xspace}
\newcommand{\kms}{\ensuremath{\mathrm{km\,s^{-1}}}\xspace}

\title[Water and clouds on WASP-166~b]{The hot Neptune WASP-166~b with ESPRESSO III: A blue-shifted tentative water signal constrains the presence of clouds}

\author[M. Lafarga et al.]{
M.~Lafarga,$^{1,2}$\thanks{E-mail: marina.lafarga-magro@warwick.ac.uk}
M.~Brogi,$^{1,3,4}$
S.~Gandhi,$^{5,1,2}$
H.~M.~Cegla,$^{1,2}$\thanks{UKRI Future Leaders Fellow}
J.~V.~Seidel,$^{6}$
L.~Doyle,$^{1,2}$
R.~Allart,$^{7}$
\newauthor
N.~Buchschacher,$^{8}$ 
M.~Lendl,$^{8}$
C.~Lovis,$^{8}$ 
D.~Sosnowska$^{8}$ 
\\
$^{1}$Department of Physics, University of Warwick, Gibbet Hill Road, Coventry CV4 7AL, United Kingdom\\
$^{2}$Centre for Exoplanets and Habitability, University of Warwick, Coventry, CV4 7AL, UK\\
$^{3}$INAF - Osservatorio Astrofisico di Torino, Via Osservatorio 20, 10025, Pino Torinese, Italy\\
$^{4}$Dipartimento di Fisica, Universit\`a degli Studi di Torino, via Pietro Giuria 1, I-10125, Torino, Italy\\
$^{5}$Leiden Observatory, Leiden University, Postbus 9513 2300 RA, Leiden, The Netherlands\\
$^{6}$European Southern Observatory, Alonso de Córdova 3107, Vitacura, Regi\'on Metropolitana, Chile\\
$^{7}$Department of Physics, and Trottier Institute for Research on Exoplanets, Universit\'e de Montr\'eal, Montr\'eal, H3T 1J4, Canada \\
$^{8}$Observatoire Astronomique de l'Universit\'e de Gen\`eve, Chemin Pegasi 51b, CH-1290 Versoix, Switzerland \\
}

\date{Accepted XXX. Received YYY; in original form ZZZ}

\pubyear{2022}

\usepackage{newtxtext,newtxmath}

\begin{document}
\label{firstpage}
\pagerange{\pageref{firstpage}--\pageref{lastpage}}
\maketitle

\begin{abstract}
% This is a simple template for authors to write new MNRAS papers.
% The abstract should briefly describe the aims, methods, and main results of the paper.
% It should be a single paragraph not more than 250 words (200 words for Letters).
% No references should appear in the abstract.
With high-resolution spectroscopy we can study exoplanet atmospheres and learn about their chemical composition, temperature profiles, and presence of clouds and winds, mainly in hot, giant planets. State-of-the-art instrumentation is pushing these studies towards smaller exoplanets. Of special interest are the few planets in the `Neptune desert', a lack of Neptune-size planets in close orbits around their hosts. %Studying these can provide insight to their planetary formation and evolution, and shed light on the origin of the desert. 
Here, we assess the presence of water in one such planet, the bloated super-Neptune WASP-166~b, which orbits an F9-type star in a short orbit of 5.4 days. Despite its close-in orbit, WASP-166~b preserved its atmosphere, making it a benchmark target for exoplanet atmosphere studies in the desert. We analyse two transits observed in the visible with ESPRESSO. We clean the spectra from the Earth's telluric absorption via principal component analysis, which is crucial to the search for water in exoplanets. We use a cross-correlation-to-likelihood mapping to simultaneously estimate limits on the abundance of water and the altitude of a cloud layer, which points towards a low water abundance and/or high clouds. 
We tentatively detect a water signal blue-shifted $\sim$5~\kms from the planetary rest frame.
% Finally, we perform injection tests which reveal that, if present, a water signal would have been detected with high significance.
Injection and retrieval of model spectra show that a solar-composition, cloud-free atmosphere would be detected at high significance. This is only possible in the visible due to the capabilities of ESPRESSO and the collecting power of the VLT. 
This work provides further insight on the Neptune desert planet WASP-166 b, which will be observed with \emph{JWST}.
\end{abstract}


%%%%%%%%%%%%%%%%%%%%%%%%%%%%%%%%%%%%%%%%%%%%%%%%%%

%%%%%%%%%%%%%%%%% BODY OF PAPER %%%%%%%%%%%%%%%%%%

\section{Introduction}

High-resolution spectroscopy is used to detect and characterise the atmospheres of transiting planets, giving us information about their chemical composition, temperature profiles, and the presence of clouds and winds, mainly in hot, giant planets \citep[see e.g.][for a review]{birkby2018review}. State-of-the-art instrumentation is pushing the precision of our measurements towards the detection and characterisation of the atmospheres of cooler and smaller exoplanets (Neptune and Earth-sized planets).
% Water
One of the best studied chemical species with high-resolution instruments is water. 
Water analyses have mainly been focused in the infrared wavelength range, because its spectrum presents several strong absorption bands, while in the optical range, there are only few weaker absorption bands in the red.
% The study of water vapour with high-resolution spectroscopy has been mostly focused on the near-infrared wavelength range.
% The spectrum of water vapour presents several strong absorption bands in the infrared, and a few weaker absorption bands in the red part of the visible wavelength range.
Water vapour has been targeted by several ground-based, high-resolution infrared spectrographs such as CRIRES \citep{kaeufl2004CRIRES}, NIRSPEC \citep{mclean1998NIRSPEC}, GIANO \citep{origlia2014GIANO}, CARMENES NIR \citep{quirrenbach2016carmenes}, and SPIRou \citep{donati2020spirou}. Observations with these instruments have led to the detection of water vapour in the atmospheres of several transiting and non-transiting exoplanets \citep[e.g.][]{birkby2013HD189733water,brogi2014HD179949COwater,brogi2016rotationwinds,piskorz2017UpsAndWater,birkby2017water51Peg,brogi2018gianoHD189733,alonsofloriano2019HD189733water,sanchezlopez2019HD209458water,webb2020HD179949water,boucher2021HD189733spirouWater,webb2022TauBooWater}.
However, detections of water in the visible range remain challenging. %, mainly due to the fact that the absorption bands present in the visible are weaker than those at longer wavelengths. and clouds

% Water in the visible
\citet{esteves2017water55Cane} and \citet{jindal2020hires55Cane} studied the presence of water in the super-Earth 55~Cancri~e with several transits obtained with the optical, high-resolution spectrographs HDS \citep[wavelength range 5240-7890~\AA,][]{noguchi2002HDS} on the 8.2~m Subaru telescope, ESPaDOnS \citep[5060-7950~\AA,][]{donati2003espadons} on the 3.6~m CFHT, and GRACES \citep[3990-10480~\AA,][]{chene2014GRACES} on the 8.1~m Gemini North telescope. They did not detect water and ruled out the presence of water-rich atmospheres if cloud-free.
\citet{deibert2019saturns} studied HDS and GRACES observations of HAT-P-12~b and WASP-69~b, two warm sub-Saturns with inflated radii. They also did not detect water, but injection tests suggest a cloudy atmosphere with a small amount of absorption, in agreement with other studies.

\citet{allart2017waterHD189733} used HARPS \citep[3780-6910~\AA,][]{mayor2003harps} on ESO's La Silla 3.6~m telescope to look for water in the gas giant HD~189733~b, focusing on the 6500~\AA~absorption band. The data used is too noisy to constrain the presence of water and the authors estimated that over 10 HARPS transits would be needed to have a constrain at a significant level. However, they also estimated that a significant detection would be feasible with only a single ESPRESSO transit, due to its increased collecting power and the fact that its wavelength range includes a stronger water band at $\sim$7400~\AA.

With ESPRESSO \citep[3782-7887~\AA,][]{pepe2021espresso} on ESO's 8.2~m VLT, \citet{allart2020wasp127} studied WASP-127~b, a super-Neptune-mass planet with a radius larger than that of Jupiter, which makes it an extremely bloated planet. No water was found but, together with low-resolution data, the authors were able to constrain the pressure of a cloud-deck.
Also with ESPRESSO, \citet{sedaghati2021wasp19espresso} observed the hot Jupiter WASP-19~b, which orbits a G8~V star in less than 1~day. Water was again not detected, but in this case, injection tests showed that it would only be detectable at high abundances and not feasible with ESPRESSO on an 8-m class telescope. The authors argued that this is the case due to the relatively faint host star ($V=12.3$~mag) and the short transit duration \citep[1.6~hours,][]{corteszuleta2020tramos}, which result in few in-transit observations with low signal-to-noise ratio (S/N).

Finally, \citet{sanchezlopez2020waterCarmvis} reported a water detection in one out of three transits of HD~209458~b using the 7000 to 9600~\AA\,absorption bands present in the red part of the visible arm of CARMENES VIS \citep[5200-9600~\AA,][]{quirrenbach2016carmenes}. Injection tests indicated that the lack of detection in the other two nights could be due to a lower S/N and a higher degree of telluric variability, which results in a worse telluric removal that hinders the detection of water.

% Neptune desert
% Of special interest are the few planets located in or close to the `Neptune desert', a lack of Neptune-size planets in close orbits around their host stars. The study of such planets can provide insight to their formation and evolution, and the existence of the desert.\todo{Expand and add references.}
% "Formation of clouds is an issue of increasing importance as the atmosphere becomes cooler."
% 
% However, the atmospheres of these planets have proven challenging to characterise due to the presence of high-altitude aerosols such as clouds and hazes. Clouds and hazes reduce the strength of the features observed in an exoplanet spectrum, affecting the detectability of species such as water.
% Low-resolution observations of several low-temperature planets show featureless spectra, which have been attributed to the presence of thick high-altitude clouds.
% Clouds and hazes do not only affect cool planets, but their presence has also been inferred in several hot Jupiters, which show mutted water features.

% Clouds
As seen from the previous results, an important feature observed in the atmospheres of both hot and cool planets is the presence of clouds and/or hazes. Clouds and hazes reduce the strength of the features observed in an exoplanet spectrum, affecting the detectability of species such as water.
Low-resolution observations of planets over a range of temperatures have shown muted water spectral features compared to what is expected for cloud-free atmospheres with solar metallicity \citep[e.g.][]{sing2016hotJupiters,stevenson2016HAT-P-26,barstow2017hotJupiters,wakeford2017HAT-P-26,wakeford2017WASP-101,wakeford2019forecast,pinhas2019hotJupiters,benneke2019GJ3470,benneke2019K2-18,kreidberg2020HD106315}, and some low-mass, low-temperature planets even show completely featureless spectra \citep{kreidberg2014GJ1214,knutson2014GJ436,knutson2014HD97658b}. 
These muted features can be attributed to either the presence of thick, high-altitude clouds, or to inherently low water abundances.

As opposed to low-resolution observations, which are sensitive to broad-band spectral features, high-resolution spectroscopy is able to resolve individual lines. The cores of absorption lines are formed higher up in the atmosphere than their wings. Therefore, high-resolution data is sensitive to high-altitude regions of the atmosphere and can probe above clouds.

The abundance of the species present in exoplanet atmospheres has been typically derived from low-resolution spectroscopy, which is sensitive to broad-band features over the continuum. Opposite to that, high-resolution observations do not preserve that continuum flux needed to measure abundances. However, it has recently been shown that by using a Bayesian framework, it is possible to recover abundances from the line-to-line and line-to-continuum contrast ratio alone \citep{brogiline2019cc,gibson2020logL,line2021wasp77A,pelletier2021tauboo}.
Therefore, high-resolution spectroscopy is both sensitive to clouds and water abundance, and can break degeneracy between the two \citep{gandhi2020clouds,hood2020hazy}.

% WASP-166
In this work, we use optical, high-resolution spectroscopy to study the presence of water and clouds on the transiting planet WASP-166~b, a bloated super-Neptune. WASP-166~b orbits a relatively bright ($V=9.36$~mag), F9-type star in a close orbit of 5.4 days, at 0.06 AU \citep[][see Table \ref{tab:sysparams} for the system parameters adopted here]{hellier2019wasp166,bryant2020wasp166}. The planet has a mass of $0.101\pm0.005~\Mjup$ (1.9~\Mnep) and a radius of $0.63\pm0.03~\Rjup$ (1.8~\Rnep) \citep{hellier2019wasp166}, and its orbit has been found to be aligned with the stellar spin \citep{hellier2019wasp166,doyle2022wasp166,kunovachodzic2022rrm}.
It is located in the so-called `Neptune desert', a dearth of Neptune-size planets in close orbits around their host stars. The study of such planets can provide insight to their formation and evolution, and the existence of the desert.

Despite its close-in orbit, the planet has preserved its atmosphere, making it a benchmark target for exoplanet atmosphere studies in the Neptune desert. \citet{seidel2020wasp166,seidel2022wasp166} recently confirmed the presence of sodium in the atmopshere of WASP-166~b with high-resolution ground-based transit observations obtained with the spectrographs HARPS \citep{mayor2003harps} and ESPRESSO \citep{pepe2021espresso}, respectively.
In the optical, other than sodium, we also expect the presence of potassium (although its signature usually overlaps with strong absorption from telluric oxygen, challenging its detection) and water, which we study here \citep[e.g.][]{fortney2008hotjupiters,madhusudhan2012CtoO,moses2013hotneptunes,woitke2018eqchem,drummond2019CtoO}. Other species with signatures in the optical such as CH$_4$ or NH$_3$ do not have reliable opacities below 0.5 - 1.0 micron, and hence have not been considered here. The planet is not hot enough to have other species with optical signatures such as Fe, TiO, or VO.
WASP-166 is scheduled to be observed from space in the near-infrared with \emph{JWST}, which should constrain the presence of molecules such as \water, CO, CH$_4$, CO$_2$, C$_2$H$_2$, HCN, and NH$_3$ in the planetary atmosphere \citep{mayo2021wasp-166jwst}.
% Here, we study the presence of water vapour and clouds in WASP-166 using visible (378.2 – 788.7 nm) observations obtained with the high-resolution spectrograph ESPRESSO.

In section \ref{sec:obs} we describe the ESPRESSO observations used. Section \ref{sec:methods} details the analysis performed, and in Section \ref{sec:results} we show and discuss the results obtained. We summarise our findings and conclude in Section \ref{sec:conclusion}.

%%%%%%%%%%%%%%%%%%%%%%%%%%%%%%%%%%%%%%%%%%%%%%%%%%
%%%%%%%%%%%%%%%%%%%%%%%%%%%%%%%%%%%%%%%%%%%%%%%%%%

\section{Observations}\label{sec:obs}

We observed two full transits of WASP-166~b, on 31st December 2020 and 18th February 2021, with the high-resolution, optical (wavelength range 3782 -- 7887~\AA) spectrograph ESPRESSO \citep[][]{pepe2021espresso} installed on the VLT at the ESO Paranal Observatory, in Chile (ESO programme ID: 106.21EM, PI: H. M. Cegla).
The observations were carried out in the 1-UT configuration (using UT1 on the first night and UT4 on the second) and high-resolution mode with $2\times1$ readout binning (HR21 mode, median resolving power of $R=138\,000$). The target was observed with fibre A while fibre B was used to monitor the sky (i.e. simultaneous sky mode).
% Reduction
% Header: HIERARCH ESO PRO REC1 PIPE ID = 'espdr/2.3.1' / Pipeline (unique) identifier  
The observations were reduced with the ESPRESSO Data Reduction Software\footnote{\href{www.eso.org/sci/software/pipelines/espresso/ espresso-pipe-recipes.html}{www.eso.org/sci/software/pipelines/espresso/ espresso-pipe-recipes.html}} (DRS) version 2.3.1, which performs standard reduction steps for echelle spectra, including bias and dark subtraction, optimal order (2D spectra) extraction, bad pixel correction, flat-fielding and de-blazing, wavelength calibration, as well as extraction of sky spectrum from fibre B \citep[see][for details]{pepe2021espresso}.
In our analysis, we used the blaze-corrected and sky-subtracted (i.e. corrected for telluric sky emission) 2D spectra.

The observations of each night cover the full planetary transit \citep[transit duration 3.608~h,][]{doyle2022wasp166} and 2 to 3~hours of out-of-transit baseline (in total, before and after the transit).
The exposure time was set to 100~s to ensure a S/N sufficiently high to have photon-noise dominated spectra (S/N$\sim50$ at 550~nm) and to obtain a good temporal cadence to sample the transit. These observations were initially obtained to perform a study of the Rossiter-McLaughlin effect, which requires a fine temporal cadence during the transit \citep[see][]{doyle2022wasp166}.
In the first night, we obtained 80 in-transit observations and 26/40 observations before/after the transit, and in the second night, 81 in-transit observations and 30/27 out-of-transit observations before/after the transit.
% transit duration: $3.608^{+0.020}_{-0.015}$~h
% 1st night: 26 pre oot, 80 it, 40 post oot
% 2nd night: 30 pre oot, 81 it, 27 post oot

\begin{figure*}
\centering
\includegraphics[width=\textwidth]{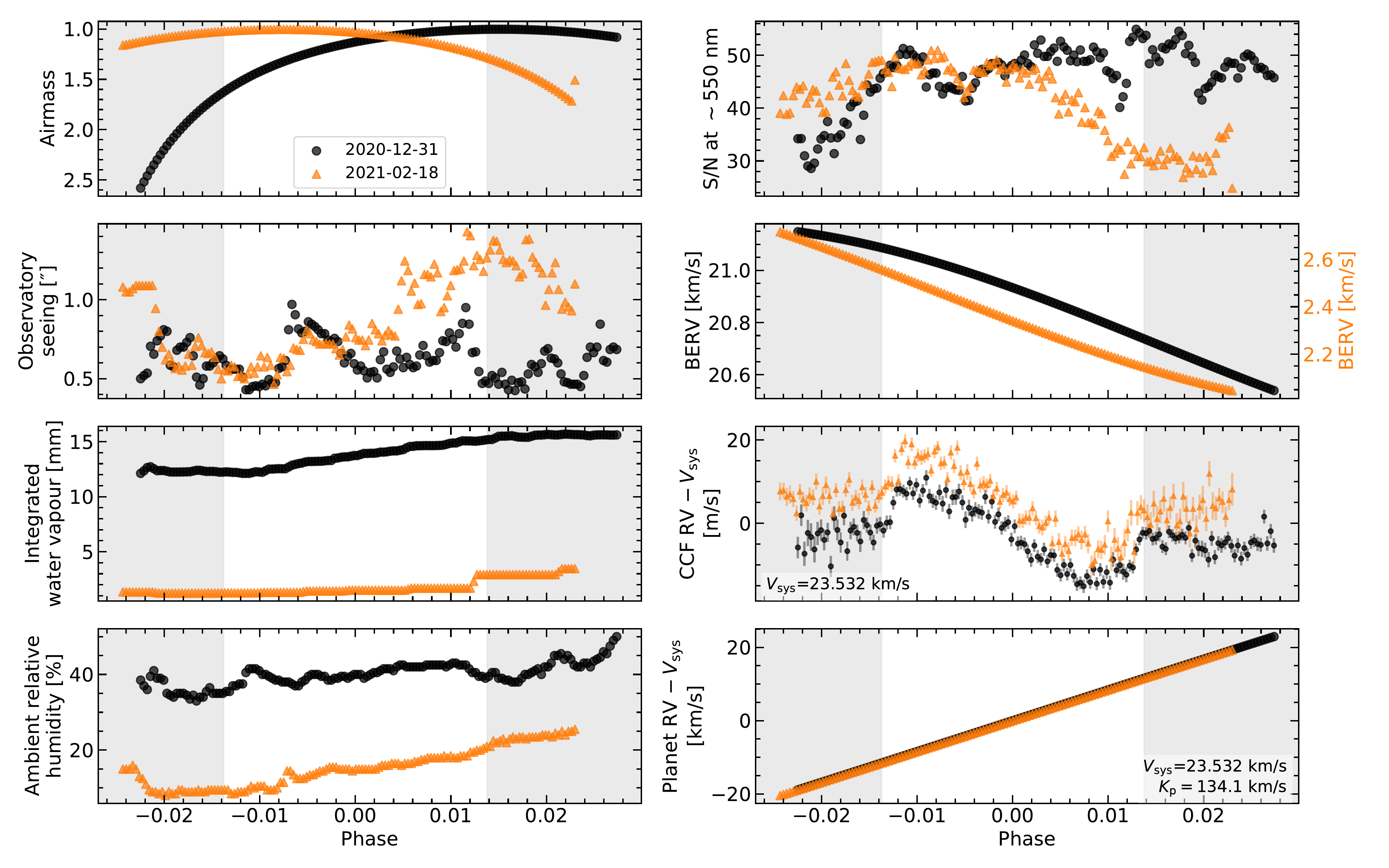}
\caption{Observing conditions, S/N, and Earth, star, and planet RVs for the 2 transits observed, as a function of the planetary phase, where phase 0 corresponds to the mid-transit. Left, top to bottom panels: Airmass (mean between start and end of each observation), seeing (mean between start and end of each observation), integrated water vapour (mean between start and end of each observation), and ambient humidity.
Right, top to bottom panels: S/N at $\sim$550~nm, barycentric Earth radial velocity (BERV, note the offset between nights), star RV from the DRS CCF, and planet RV. The $V_\mathrm{sys}$ of the system has been subtracted from the star and planet RVs (see Table \ref{tab:sysparams}). All parameters obtained from the observations FITS headers, except for the planet RV, which is computed from the orbital parameters of the system (see text).
Grey areas indicate out-of-transit phases.}
\label{fig:dataobs}
\end{figure*}

For the two nights, most of the observations were taken at low airmass ($<1.5$, see Figure \ref{fig:dataobs} for an overview of the observing conditions).
We discarded the first 8 observations of the first night (all out-of-transit observations) because they were taken at an airmass larger than 2.2, which is the maximum value for which the ESPRESSO ADC (Atmospheric Dispersion Corrector) is calibrated for.
Additionally, in the second night, we discarded 3 observations taken during the post-transit baseline due to telescope vignetting.

% The seeing remains relatively stable during the two transits, and only increases to values $>1$ during the last part of the second night, due to passing thin clouds. This is reflected in the S/N, which is between 40 -- 50 at $\sim$550~nm, and drops to below 30 in the last part of the second night.
% During the first transit, the S/N increases from 30 to 40 at the beginning due to decreasing airmass, and remains within 40 and 50 during the rest of the night.
% % 
% The integrated water vapour and the ambient humidity of both nights slightly increase from the start to the end of the observations, and both values are larger for the first night.
% % An overview of the observing conditions can be found in Figure \ref{fig:dataobs}.

We note that in the stellar RVs there is an offset of about 10~\ms in the systemic velocities of the two nights. These stellar RVs are obtained with the ESPRESSO DRS by computing the cross-correlation function with a suitable stellar mask. The reason for this offset is unknown but we attribute it to instrumental effects or differences in the observing conditions between the two nights. Regardless of the origin, the offset is too small to have any effect on our analysis (our precision is of about 1~\kms, 100 times larger than the offset). In the following, we consider as the systemic velocity of the system the average of the systemic velocities of each night.

The same observations have been used in \citet{doyle2022wasp166} to study the Rossiter-McLaughlin effect, characterise centre-to-limb convection-induced variations, and refine the star-planet obliquity, and in \citet{seidel2022wasp166} to detect the presence of sodium in the planetary atmosphere.

% ++++++++++++++++++++++++++++++++++++++++++++++++++++++++
% Obs data
% ++++++++++++++++++++++++++++++++++++++++++++++++++++++++

% ++++++++++++++++++++++++++++++++++++++++++++++++++++++++
% Table system parameters
\begin{table}
\centering
\caption{WASP-166 system properties used in this work.}
\begin{tabular}{lcl}
\hline 
Parameter & Value & Reference\\
\hline
$R_{\mathrm{p}} / R_\star$       & $0.05177^{+0.00063}_{-0.00035}$ & \citet{doyle2022wasp166} \\
$R_{\mathrm{p}}~[R_\mathrm{J}]$ & $0.6155^{+0.0306}_{-0.0307}$ & \citet{doyle2022wasp166} \\
% $R_\star~[R_\odot]$             & $1.22\pm0.06$ & \citet{doyle2022wasp166} \todo{comp. up down} \\
$a/R_\star$                      & $11.83^{+0.29}_{-0.68}$ & \citet{doyle2022wasp166}\\
$a$ [AU]                         & $0.0668^{+0.0040}_{-0.0044}$ & \citet{doyle2022wasp166}\\
$i_\mathrm{p}$ [$^{\circ}$]      & $88.85^{+0.74}_{-0.94}$ & \citet{doyle2022wasp166}\\
$t_\mathrm{0}$ [BJD]             & $2458524.40869201^{+0.00030021}_{-0.00029559}$ & \citet{doyle2022wasp166}\\
$T_\mathrm{dur}$ [hours]         & $3.608^{+0.020}_{-0.015}$ & \citet{doyle2022wasp166}\\
$P$ [days]                       & $5.44354215^{+0.00000307}_{-0.00000297}$ & \citet{doyle2022wasp166}\\
$e$                              & 0.0 & \citet{hellier2019wasp166}\\
\Vsys [\kms] & $23.532\pm 0.012$ & \citet{doyle2022wasp166}\\
\Kp [\kms] & $134.1\pm8.1$ & This work\\
$T_{\mathrm{eff}}$ [K]           & 6050$\pm$50 & \citet{hellier2019wasp166}\\
$T_{\mathrm{eq}}$ [K]           & 1270$\pm$30 & \citet{hellier2019wasp166}\\
\hline 
\end{tabular}
\vspace{2mm}
\begin{flushleft}
{\bf Notes:} Values from \citet{doyle2022wasp166} have been derived using the same ESPRESSO observations as here, as well as \emph{TESS} and NGTS photometry. In particular, \Vsys has been measured from the out-of-transit cross-correlation functions of the ESPRESSO data and here we use the mean \Vsys of the two nights \citep[][see their Table 1]{doyle2022wasp166}. \Kp has been computed here based on the parameters from \citet{doyle2022wasp166} (see Section \ref{sec:cc} and Appendix \ref{sec:ccdetails}).
\end{flushleft}
\label{tab:sysparams}
\end{table}
% ++++++++++++++++++++++++++++++++++++++++++++++++++++++++

%%%%%%%%%%%%%%%%%%%%%%%%%%%%%%%%%%%%%%%%%%%%%%%%%%
%%%%%%%%%%%%%%%%%%%%%%%%%%%%%%%%%%%%%%%%%%%%%%%%%%

\section{Methods}\label{sec:methods}

%%%%%%%%%%%%%%%%%%%%%%%%%%%%%%%%%%%%%%%%%%%%%%%%%%

\subsection{Telluric correction: PCA}\label{sec:pca}

\begin{figure*}
\centering
\includegraphics[width=\textwidth]{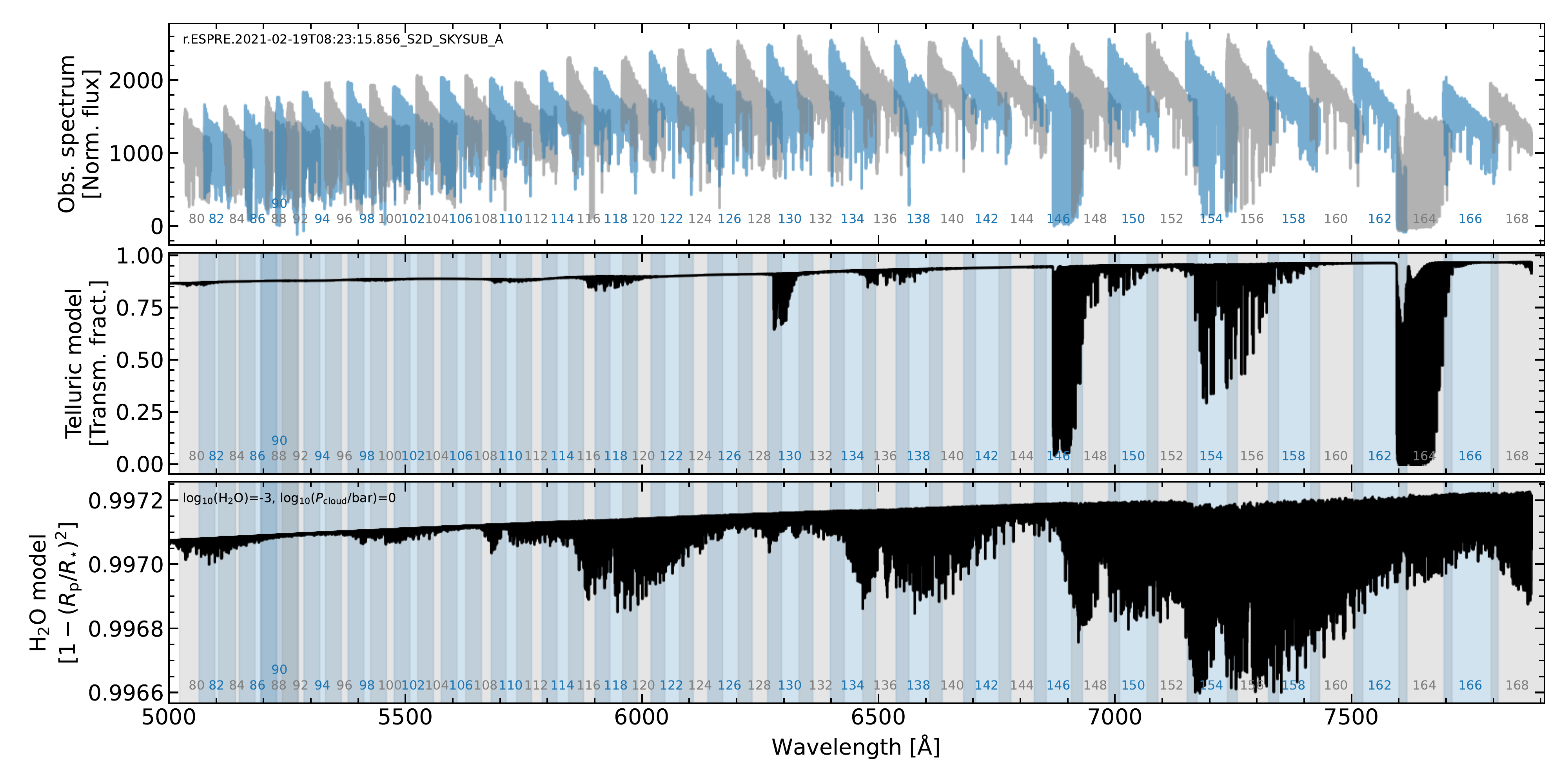}
\caption{Example of an observed spectrum of WASP-166 (top), telluric template used to select telluric-affected regions (middle), and one of the \water models used to compute the CC functions (bottom). 
Grey and blue shaded regions and numbers indicate the different ESPRESSO orders (also colour-coded in the WASP-166 spectrum in the top panel). The numbers correspond to the ESPRESSO slices (each order has two slices), starting at 0 for the first slice of the bluest order). We only show the wavelength range covering the spectral region used (where planetary model shows stronger absorption.}
\label{fig:ord_tpl_model}
\end{figure*}

Spectroscopic observations taken from the ground are affected by spectral features produced by the Earth's atmosphere, known as telluric contamination. The ESPRESSO wavelength range is affected mainly by water (\water) and oxygen (O$_2$), which produce absorption lines at specific wavelength ranges with varying strength, from shallow lines called microtellurics to deep and strong lines with completely saturated cores. The strength of the lines can vary depending on the observing conditions, such as the airmass or the atmospheric water vapour content.
The effect of tellurics is especially relevant when trying to study water in exoplanet atmospheres. This is because the planetary water absorption lines can overlap in wavelength space with the telluric water (see e.g. Figure \ref{fig:ord_tpl_model}). Hence, we need to correct our observed spectrum from telluric lines.

To correct for telluric effects, we used a principal component analysis (PCA) on the observed spectral time series inspired by \citet{giacobbe2021HD209458giano} \citep[see also][for other examples of works implementing PCA to study exoplanet atmospheres]{dekok2013HD189733,piskorz2016HD88133,piskorz2017UpsAnd,damiano2019PCA}. We design our own automated algorithm to select the number of PCA components (described in Section \ref{sec:ccopt}) and to only feed into the PCA the spectral channels most affected by tellurics (Section \ref{sec:tellpca}). 

The use of PCA to remove tellurics is based on the fact that, during the transit observations, the Earth and the target star remain stationary or quasi-stationary, while the target planet moves tens of \kms as it orbits around the star. Therefore, telluric and stellar spectral lines are always approximately located in the same pixels in the detector CCD, as they only experience a small shift in RV, while the planetary signal will shift noticeably in pixel space (see Figure \ref{fig:dataobs}).

% In our case, the barycentric Earth radial velocity (BERV) changes from 21.085 to 20.739~\kms during the first transit, and from 2.559 to 2.147~\kms during the second night, which represents a change of 346 and 412~\ms for each night (see Figure \ref{fig:dataobs}).
% The stellar RV change during each night due to its orbital movement is of the \ms order, and the RV anomaly due to the Rossiter-McLaughlin effect during the transit has an amplitude of $\sim$30~\ms.
% On the other hand, the planet RV changes from $-$11.382 to 11.462~\kms in the first transit and from $-$11.512 to 11.576~\kms in the second night, a change of the order of 23~\kms, way larger than the Earth and stellar RV shifts.
% 
% BERV 1st night: 21132.5974002287 (21148.4287835869 w/o discarding) -- 20539.0901169243
% BERV 1st transit: 21085.2571354885 -- 20739.5315395276
% BERV 2nd night: 2716.59701854939 -- 2046.4958856629700
% BERV 2nd transit: 2558.51398417979 -- 2146.97552650195
% Stellar RV 1st night: 23.5280070511328 (23.5261885630048 w/o discarding) -- 23.526668747373
% Stellar RV 2nd night: 23.5396349042463 -- 23.5400917396797
% but vsini is what matters
% Planet RV 1st night: -16573.584128016800 (-18876.443815934300 w/o discarding) -- 22938.019005478900
% Planet RV 1st transit: -11382.47767451080 -- 11462.039503312700
% Planet RV 2nd night: -20426.566953665400 -- 19298.45377246460
% Planet RV 2nd transit: -11512.652359096500 -- 11576.028481283500

The PCA method consists in finding an orthogonal basis for the covariance matrix of the data in which the eigenvectors (also called principal components, PC) represent the direction of decreasing variance in the data. That is, the first vector or PC of the new basis has the direction of the maximum variance in the data, the second one has the direction of the second largest variance, and so on.
Since the first PCs are the ones that describe most of the variance in the data, we can remove them to clean the data of the strongest telluric, stellar, and instrumental time-dependent variations.

In our case, the data matrix $M$ is composed of the different observations or frames as rows ($nf$) and the pixels or spectral channels as columns ($nx$). We work slice-by-slice, therefore, the steps described below are repeated for each slice, and for each night, separately.
We note here that, conversely to other echelle spectrographs, ESPRESSO uses an APSU (anamorphic pupil slicer unit) that divides each order into two slices (i.e., the two slices corresponding to a specific order cover the same wavelength range). We treat each of the slices separately.
% We work on the time series spectra order by order, therefore, the steps described below are repeated for each order, and for each night, separately.

We detail our PCA implementation in Appendix \ref{sec:pcadetails}. To briefly summarise it here, we first cleaned the spectra from flux anomalies, standarised the data matrix $M$, and then performed the PCA. Instead of directly decomposing the covariance matrix of the data as in \citet{giacobbe2021HD209458giano}, we applied the PCA via a singular value decomposition (SVD).

% ------------------------------------------------
In this work, we are only studying the presence of water in the planetary atmosphere of WASP-166~b. Therefore, we are mostly concerned in the removal of telluric lines from the observed data.
The host star, WASP-166, is too warm to display any water in the stellar spectrum \citep[spectral type F9~V and $T_\mathrm{eff}=6050$,][]{hellier2019wasp166}. The star is not especially active and we do not expect the presence of cool spots on the photosphere to be significant. Even if spots were present, their temperature contrast with the quiet photosphere is expected to be small, and hence, not sufficiently cool to display water either.
Nevertheless, we want to note here that, in transmission spectroscopy, when studying planetary species that are also present in the stellar photosphere, one needs to account for the Rossiter-McLaughlin effect and centre-to-limb variations (CLV) across the stellar disc.
This is because, during the transit, the planet occults different areas of the rotating stellar disc, which results in the in-transit stellar spectra being distorted (mainly depending on the projected stellar rotational velocity, the stellar obliquity, and the impact parameter). These distortions need to be accounted for to derive accurate and precise estimates of the planetary transmission spectra \citep[see e.g.][for more details on such effects and strategies to account and correct for them]{brogi2016rotationwinds,yan2017clv,chiavassa2019planetstar,hoeijmakers2020wasp121,casasayas-barris2021HD209458espresso,seidel2022wasp166,maguire2022wasp121}.
% ------------------------------------------------

%%%%%%%%%%%%%%%%%%%%%%%%%%%%%%%%%%%%%%%%%%%%%%%%%%

\subsubsection{Optimisation of the number of PCA components per slice}\label{sec:ccopt}

Since different orders are differently affected by tellurics, we performed a per-slice optimisation of the number of components $NC$ to be removed when applying the PCA, which we describe in this section. To perform this optimisation, we made use of the cross-correlation function (CC) of the observed spectra with a water model. We refer the reader to the following Section \ref{sec:cc} for all the details on the CC computation.

For each slice affected by tellurics, we started by removing the first 2 components in the PCA. We then computed the CC of the resulting spectra with a water model and coadded the CCs of the in-transit observations in the barycentric rest frame. Coadding in the barycentric frame maximises the presence of telluric residuals in the CC, which is what we are focusing on at this stage.

We then assessed the significance of the telluric signal by taking the value of minimum or maximum CC flux in the region $\pm$10~\kms (to cover the full telluric feature) around the mean BERV of the observations, and comparing it with the scatter (standard deviation) of the CC flux outside of this region. 
The atmospheric water vapour changes during the observations, increasing and decreasing from the overall trend dictated by the change in airmass. This causes negative and positive residuals in the processed spectra, which result in correlation and anti-correlation with the CC water template used. Therefore, when looking at the telluric signal in the CC, we considered both minima and maxima features (i.e. anti-correlation and correlation with the template).
We considered a signal at the telluric position of the CC to be significant if the minimum (or maximum) flux is below (or above) 3.5 times the standard deviation of the flux of the rest of the CC.
If the telluric signal is significant, we repeat the process but removing an additional PCA component. This goes on until the signal is not significant, or until the algorithm reaches the maximum number of components allowed. We set the maximum number of components to be removed to 15 (after removing over $\sim15$ components, injected planetary signals start to decrease in significance).

% We see that in the observation-coadded CCFs, the dominant feature at the telluric position is in general a minimum, as opposed to the individual observation CCFs, that can show a minimum or a maximum. 
Although the aforementioned CC functions of our individual observations can show a minimum or a maximum at the expected telluric position, if we coadd the CC functions of the individual observations, the dominant feature at the telluric position in our case is a minimum, i.e., anti-correlation.
This means that the observations with telluric residuals that anti-correlate with the models are more prominent than those observations with a positive correlation. The coadding of anti-correlated and correlated CCFs can result in a smearing of the overall signal. To check for that, we also computed the significance of the telluric peak in each individual observation. We observe that for all the cases where we have a significant signal in the `observation-coadded' CC function, more than half of the individual observations also show a significant signal. Additionally, if more than half of the observations contain a significant telluric signal, so does the coadded CC function.

We note again that here, instead of coadding all the available observations, we coadded only the in-transit ones. This is because these observations are the only ones we use in the planet analysis, and therefore we are mostly concerned about the telluric effects in them. Aside from this, we noticed that the observations at high airmass (airmass higher than 2 at the beginning of the first night, and airmass close to 1.7 at the end of the second night) are the ones that show the strongest telluric signals in the CC function, being very distinct than those immediately after or before.
%If we included these high airmass observations in the coadded CC function, they heavily biased the significance of the telluric signal, and made the algorithm to keep removing components. 
If we included these high airmass observations in the coadded CC function, they heavily biased the significance of the telluric signal, so that the algorithm keeps removing components even though the in-transit telluric signal is not significant.

% However, we want to consider all the observations (even the relatively high airmass ones) when applying the PCA to maximise the number of out-of-transit observations, where only the telluric signal is present, and not the planet one.
% However, we want to apply the PCA to all the observations, even those at relatively high airmasss, to maximise the number of out-of-transit observations. This is because in the out-of-transit observations, only the telluric signal is present, and not the planet one, which should help the PCA to not select the planetary signal.

%%%%%%%%%%%%%%%%%%%%%%%%%%%%%%%%%%%%%%%%%%%%%%%%%%

\subsubsection{Selectively feeding telluric lines into the PCA}\label{sec:tellpca}

To try to further improve the telluric removal, instead of using the whole spectral range of each slice, we tested feeding into the PCA only the pixels affected by tellurics, i.e. pixels containing telluric lines. By doing this, the PCA should better trace the variability due to telluric changes. % rather than other possible systematic effects present in the spectrum.

% ESO Sky Model
To determine the telluric-affected pixels, we used the ESO Sky Model Calculator\footnote{\url{https://www.eso.org/observing/etc/skycalc}} based on the Cerro Paranal Advanced Sky Model \citep{noll2012skymodel} to generate a telluric absorption model in the ESPRESSO wavelength range (see Figure \ref{fig:ord_tpl_model}, middle panel). We interpolated the model to the observed wavelength grid of each slice and continuum-normalised it by fitting a cubic spline (we do this slice by slice). To fit the spline, we selected the pixel with maximum flux in windows of 25 pixels and avoided strong telluric bands that would bias the determination of the continuum. This results in a flat telluric spectrum normalised to one.

After this normalisation, we flagged as telluric-affected all the pixels that overlap with a telluric line. We set the threshold to pixels where the telluric flux is below 0.998, which allows us to select most of the lines present in the ESPRESSO spectral range. The slices affected are 80-83, 96-103, 108-123, 128-141, 146-169 (slice numbering starts at 0 for the first slice of the bluest order). When applying the PCA, only these pixels are used in the SVD. In orders with no telluric lines, we still used all the pixels to remove any systematics.

%%%%%%%%%%%%%%%%%%%%%%%%%%%%%%%%%%%%%%%%%%%%%%%%%%

\subsection{High-resolution cross-correlation spectroscopy}\label{sec:cc}

After correcting for tellurics with the PCA, we used the high-resolution cross-correlation spectroscopy (HRCCS) method to search for the presence of water in the atmosphere of WASP-166~b. Planetary water produces thousands of molecular absorption lines in the planetary transmission spectrum. This water signal, however, is below the noise level of the data. The HRCCS method coadds all the lines present in the transmission spectrum by cross-correlating the processed observations with an adequate spectral template of the planetary atmosphere. To compute the CC, the template is Doppler-shifted by a range of RV values, and, for each shift, we take the dot product with the observed data.
% multiplied by the observed data (specifically, we perform a dot product of the model and the data, i.e., a per-element sum of the multiplied spectra). 
This operation results in a cross-correlation function with much higher S/N than a single spectral absorption line, which enhances the planetary signal. This is because the S/N of the CC function scales with the square root of the number of lines coadded when computing the CC. In Section \ref{sec:models}, we describe the different sets of CC models used, and in Sections \ref{sec:CCtologL} and \ref{sec:confidenceintervals}, we explain the formalism used to compute the CC and assess the significance of the results within the cross-correlation-to-$\log$ likelihood (CC-to-$\log L$) framework.

%%%%%%%%%%%%%%%%%%%%%%%%%%%%%%%%%%%%%%%%%%%%%%%%%%

\subsubsection{Planetary atmosphere water models}\label{sec:models}

\begin{figure*}
\centering
\includegraphics[width=\textwidth]{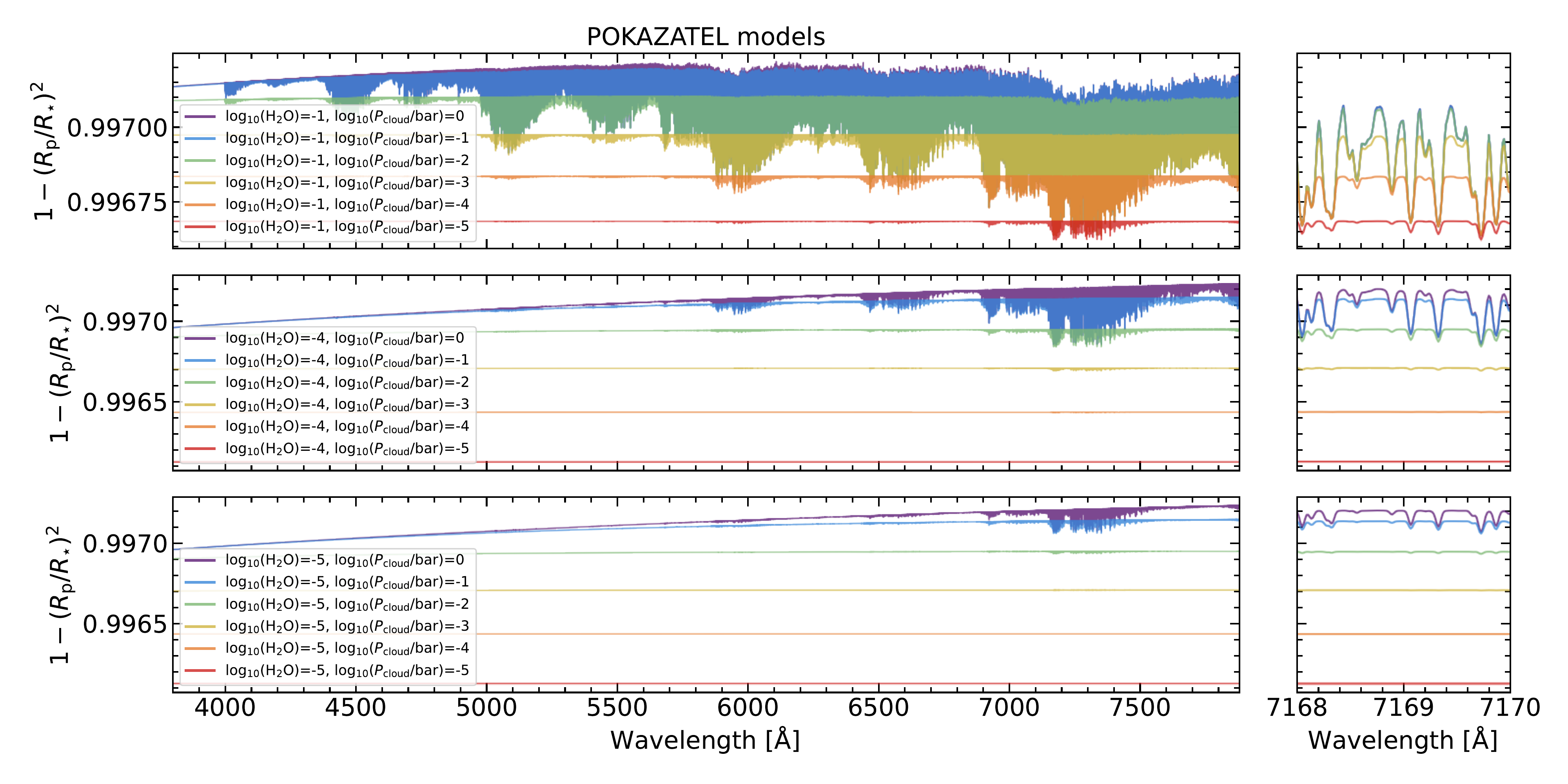}
\caption{Left: POKAZATEL \water templates for WASP-166~b covering the ESPRESSO wavelength range for a range of water abundances ($\log _{10}(\mathrm{H_2O})=-1$ top, $\log _{10}(\mathrm{H_2O})=-4$ middle, and  $\log _{10}(\mathrm{H_2O})=-5$ bottom), and a range of cloud deck pressures (depicted by various colours in all panels). Right: Zoom in on a region with strong absorption lines. This figure shows that the strength of the water absorption lines decreases as we decrease the abundance and/or decrease the cloud deck pressure.
}
\label{fig:models}
\end{figure*}

We generated primary eclipse spectra of WASP-166~b using GENESIS adapted for transmission spectroscopy \citep{gandhi2017genesis, pinhas2018aura}. 
GENESIS is a line-by-line numerical radiative transfer code that computes the transmission spectrum of the atmosphere given the atmospheric temperature and chemical abundance profile. The opacity of each species is computed on a grid of pressure-temperature ($P$-$T$) values for each wavelength to determine the overall optical depth of rays passing through the atmosphere and therefore the transit depth at each wavelength. 
We use a grid of fixed pressure values, between 100 to 10$^{-7}$ bar and evenly spaced in $\log P$. 
We assumed an isothermal temperature profile consistent with the equilibrium temperature of WASP-166~b, $\sim1270$~K. 
The chemical abundances are set as volume mixing ratios (VMR) assumed to be vertically constant throughout the atmosphere.
We also included a wavelength-independent cloud deck at different pressures by setting all wavelengths to a very high opacity.

The models spanned a grid in \water abundance and cloud pressure, encompassing $\log _{10}(\mathrm{H_2O})=-1$ (highest abundance, in VMR) to $-5$ (lowest abundance), and cloud deck pressures of $\log _{10}(P_\mathrm{cloud}\mathrm{/bar})=0$ (lowest altitude) to $-5$ (highest altitude), both in steps of 0.5~dex (see Figure \ref{fig:models} for examples). 
% These were generated with an \textbf{isothermal} temperature profile consistent with the equilibrium temperature of WASP-166~b, $\sim1270$~K. 
In total, we computed two grids of model spectra, one using an ExoMol POKAZATEL \citep{polyansky2018pokazatel} line list and the other with a HITEMP \citep{rothman2010HITEMP} line list \citep[see][for further details on opacities]{gandhi2020crosssections}. In addition, all models across both grids include collisionally-induced absorption from H$_2$-H$_2$ and H$_2$-He interactions \citep{richard2012cia} and Rayleigh scattering due to H$_2$. Each model was generated at a spectral resolution of R=500\,000 between 0.38-0.8~$\mu$m.

The models already include intrinsic pressure and temperature broadening. To better match the line shape of the expected observed planetary signal, we further broadened these model spectra by the instrument profile of the observations; for this, we used a Gaussian kernel with FWHM corresponding to the R=140,000 resolution of ESPRESSO (of $\sim2.14~\kms$).
We also computed the broadening due to planetary rotation (assuming it is tidally locked), which is of only 0.58~\kms. This is negligible compared to the instrument profile broadening, and hence, we do not consider it here (i.e. including it would only change the broadening from $\sim2.14~\kms$ to $\sim2.22~\kms$).
% To better match the line shape of the expected planetary signal, we broadened the models by the instrument profile. To do that, we convolved them with a Gaussian kernel with FWHM corresponding to the resolution of ESPRESSO.

We used two different line lists because published water lines in the optical have not been extensively empirically verified. In the optical, water absorption bands are weaker than in the near-infrared. Due to this reduced strength, the accuracy and completeness of the model lines in the optical is expected to be worse than in the near-infrared, because their experimental verification is more challenging. Therefore, there could be differences between different line lists. To check for systematics due to these potential differences we decided to repeat our analysis using the two sets of line lists.

%%%%%%%%%%%%%%%%%%%%%%%%%%%%%%%%%%%%%%%%%%%%%%%%%%

\subsubsection{CC-to-$\log L$ framework}\label{sec:CCtologL}

To assess the significance of any planetary signals, we followed the cross-correlation to $\log$-likelihood framework \citep[CC-to-$\log L$,][]{zucker2003cc2logL,brogiline2019cc,gibson2020logL}. This is a Bayesian framework based on mapping the cross-correlation function to a $\log$ likelihood function. This allows us to accurately assess the significance of any detections by deriving confidence intervals, as well as to compare the performance of different models.

We used the CC-to-$\log L$ mapping proposed by \citet{brogiline2019cc}
\begin{equation}\label{eq:cc2logL}
\log(L) = - \frac{N}{2} \log [ s^2_f - 2R + s^2_g ],
\end{equation}
where $s^2_f$ is the variance of the observed spectrum, $s^2_g$ is the variance of the model used, $R$ is the cross-covariance between the observed spectrum and model, and $N$ is the number of points in the spectrum. The cross-correlation is contained in the above equation, since the correlation coefficient $C$ is proportional to the cross-covariance $R$ as
\begin{equation}\label{eq:r2cc}
\mathrm{C} = \frac{R}{\sqrt{s^2_f s^2_g}}.
\end{equation}

In our case, all these refer to each individual spectral slice, because we are working slice-by-slice. The broadened models are sliced so that they are within the wavelength range of each order. We also spline-interpolated the models to the wavelength grid of each order, so that the number of data points $N$ of the observed spectrum and model are the same. This interpolation is performed for every RV shift for which we compute the CC and \logL functions.

We followed two different approaches to compute the CC and $\log L$ functions.
Both methods lead to the same final result but have different advantages and drawbacks, as we describe in the following paragraphs. For more details on the implementation of each approach, we refer the reader to Appendix \ref{sec:ccdetails}.
% Mainly, the implementation of the first approach is significantly faster than the second one, and it allows us to study the behaviour of the telluric signal directly in the CC, which is not possible with the second approach.
% The second one, however, is more precise because it does not require us to interpolate the \logL to different RV grids, which is needed in the first one, and because it allows us to process the model used through the same PCA as the data. This last step is needed to avoid biases when computing the CC between data and model.
% In the following sections we describe both approaches, as well as comment on their main differences.

% --- 2 methods summary ---
In our initial or `fast' approach, we compute the CC and \logL functions of each slice for a fixed RV grid. Then, for each observation, the \logL function of all the slices considered are coadded. Finally, the \logL functions of the in-transit observations are coadded in time along the planet RV, as a function of \Kp, from which we can then build the usual \KpVsys (or \KpVrest if \Vsys has been subtract) maps.
In the second or `slow' approach, instead of computing the full CC and \logL functions for a fixed grid of RV values, we only shift the model once to the expected planet RV (which is given by a pair of \Kp and \Vsys values), and compute a single CC and \logL value. We repeat this for a range of \Kp and \Vsys pairs, which also results in the usual \KpVsys maps.
%
% This is equivalent to the procedure followed in the first approach, in which we coadded the full \logL functions of the different orders of each observation, and then coadded the different observations together after being shifted to the planet rest frame (for each $K_\mathrm{p}$ considered).
With the slow approach, we are building the \KpVsys map pixel-by-pixel, while in the fast one, we directly get a full row of the map for each \Kp considered.

The main advantage of the slow approach is that it allows us to process the model used in the cross-correlation through the same PCA as the data, which is not possible in the fast approach (see Appendix \ref{sec:ccdetails} for details about the model processing). This is important because the PCA might alter the planetary signal contained in the data by changing the line strength and shape. The models used in the fast approach do not contain any change due to the PCA, therefore, the match with a possible planetary signal will not be as good as if the model has also been altered in the same way as the data. Due to this mismatch, a potential planetary detection might be weaker and biased in \Kp and \Vsys. Moreover, when performing the model comparison (see below Section \ref{sec:modelcomparison}), we could also misinterpret the water abundance and cloud deck pressure because the line depths do not match between model and data.
Therefore, we expect more accurate results with the slow approach than with the fast one, because 1) we are computing the \logL value at the exact \KpVsys, and not shifting and interpolating the whole function computed for a different \Kp and \Vsys pair, and 2) we are processing the model in the same way as the PCA modifies the data, which should result in a better match between model and data.
However, the implementation of the fast approach is significantly faster than the slow one. Moreover, the fast approach allows us to study the behaviour of the telluric signal directly in the CC and \logL functions, which is not possible with the slow approach.
In the following, we refer to the two approaches as `fast/unprocessed-model' and `slow/processed-model'.

%%%%%%%%%%%%%%%%%%%%%%%%%%%%%%%%%%%%%%%%%%%%%%%%%%

% \paragraph{Confidence intervals}
\subsubsection{Confidence intervals}\label{sec:confidenceintervals}

The CC-to-$\log L$ framework allows us to estimate confidence intervals for the \KpVsys maps \citep{brogiline2019cc,pino2020kelt-9Fe}, to know which \KpVsys pair is more likely compared to all the pairs tested.
According to Wilks' theorem \citep{wilks1938logL}, minus twice the difference between the \logL values of two models ($\Delta\log L = -2 (\log L_1 - \log L_2)$) follows a $\chi^2$ distribution with number of degrees of freedom equal to the number of explored parameters.
In our case, we are comparing the \logL value of each \KpVsys pair (2 parameters) with the maximum \logL of the map. I.e., we subtracted each \logL value from the maximum \logL of the map, which, if detected, should correspond to the planetary signal.
We can then compute the p-value of this $\chi^2$ distribution, from which we can finally derive the confidence interval value in units of standard deviation ($\sigma$) for a Normal distribution. Then, the model with the highest \logL will have a $\sigma$ of 0, with less likely models having increasing $\sigma$ values.
% dlogl = 2. * (np.nanmax(logl) - logl)
% p_value = stats.chi2.sf(dlogl, dof)  # survival function: probability to measure a chi2 that is equal to or greater than dlogL for one degree of freedom (the p-value)
% sigma_value = stats.norm.isf(p_value/2.)  # two-tail test of a normal distribution to invert this p-value and obtain the number of standard deviations you are away from the mean, which is the definition of sigma

We computed the confidence intervals for the data of each night separately and also on both nights combined. To combine the nights, we summed the \logL values of each \KpVsys pair of both nights, and then computed the confidence intervals on this coadded \logL.

%%%%%%%%%%%%%%%%%%%%%%%%%%%%%%%%%%%%%%%%%%%%%%%%%%

\subsection{Model comparison}\label{sec:modelcomparison}

We also performed a likelihood ratio test for each of the 2 (POKAZATEL and HITEMP line lists) grids of 99 models (9 water abundances $\times$ 11 cloud top pressures) computed (see Section \ref{sec:models}). This allows us to derive confidence intervals in both cloud top pressure and water abundance.
We computed the \KpVrest map (we have subtracted the expected \Vsys) for each of the 99 models. To compute the \logL functions we followed the CC approach 2 as explained in Appendix \ref{sec:ccslow}, in which we modify the template in the same way as the PCA processes the data. We used a grid ranging from 90 to 150~\kms in \Kp, in steps of 3~\kms, and from $-20$ to 5~\kms in \Vrest, in steps of 1~\kms. This grid results in a reasonable computational time, is sufficiently fine to resolve any signals, and covers the expected planetary position as well as any tentative detections seen in our initial tests.
% To compare any differences between the two CC approaches, we also repeated the analysis using the fast approach. In this case, we used a reduced RV grid to compute the \logL compared to the initial tests (which was from $-100$ to 100 in steps of 0.5~\kms), from $-25$ to 75~\kms, also in steps of 0.5~\kms. When coadding the \logL functions in planet rest frame, this covers a \Vsys from $\sim -25$ to $\sim 28$~\kms. We tested \Kp values from 0 to 250~\kms since this does not require significant extra computational time.

To identify the model with the highest significance, we compared their \logL values, following the same idea as when computing confidence intervals for the different \KpVsys pairs explained above.
Now, we have again two parameters: the water abundance $\log _{10}(\mathrm{H_2O})$ and the cloud deck pressure $\log _{10}(P_\mathrm{cloud}\mathrm{/bar})$.
In the \KpVrest map obtained for each model, we computed the maximum \logL of an area around where the planet is expected. We used an area spanning $\pm10~\kms$ from the expected \Kp, and $-10$ and $+$5~\kms from the expected \Vrest (see Table \ref{tab:sysparams}). We used $+$5~\kms in \Vrest instead of $+$10~\kms (i.e. which would be symmetric around the expected \Vrest) because we are limited by the \Vrest range covered. We tested smaller and larger areas (from $\pm$5~\kms up to $\pm$20~\kms in both \Kp and \Vrest) and the results obtained do not change significantly. 
This gives us a \logL$_\mathrm{max}$ for each model. We can then apply Wilks' theorem to obtain confidence intervals for the grid of models. As before, we computed twice the difference of the \logL$_\mathrm{max}$ of each model from the absolute maximum of all models, derived the p-value from this distribution of $\Delta$\logL, and finally computed the confidence intervals in $\sigma$.
This likelihood ratio test informs us about how likely the 99 models tested are compared to each other. The best model will then have a $\sigma$ of 0, and the rest of the models will have larger values of $\sigma$.
We performed this analysis on each night individually, as well as on both nights coadded (for which we used the \KpVrest \logL map obtained by summing the \logL values of each night).

%%%%%%%%%%%%%%%%%%%%%%%%%%%%%%%%%%%%%%%%%%%%%%%%%%

\subsection{Injection tests}

We also tested the detectability of the water signal in our data by performing several injection tests using the \water models described in Section \ref{sec:models}. We note that we do not use these injection tests to optimise our data analysis, but rather to assess the sensitivity of our data to a water signal.
We tested different strengths of the model by scaling it to different values. To scale the model, we subtracted the mean of the model flux, which removes the average transit depth and leaves only the effect of the planet atmosphere. We then multiplied the residual spectrum (which is now only due to the planet atmosphere) by a scaling factor. We then brought back the original flux level by adding the original mean.
We note that this scaling factor does not correspond to an increase or decrease in the \water abundance of the model. Increasing the abundance could lead strong lines to saturate, while this will not happen with a scaling factor. By using the scaling factor we only want to study its detectability.

Right before applying the PCA, we injected the scaled model to the in-transit observations. We performed this process slice-by-slice.
To do this, we first Doppler-shifted the wavelength of the model by the corresponding RV of the planet at the time of each observation, so that the model shift reflect that of the actual planet, and interpolated the shifted model to the wavelength grid of each observation. We then multiplied the flux of each observation by the flux of the corresponding model. This way, each in-transit observation includes now a model of the planetary spectrum. We performed this step after the observed flux had been cleaned of bad pixels. After the injection, the standarisation and the PCA are performed as explained above (see Section \ref{sec:pca}).
We then computed the CC as explained above with the same model as injected.

%%%%%%%%%%%%%%%%%%%%%%%%%%%%%%%%%%%%%%%%%%%%%%%%%%
%%%%%%%%%%%%%%%%%%%%%%%%%%%%%%%%%%%%%%%%%%%%%%%%%%

\section{Results and discussion}\label{sec:results}

In this section, we first present the results from the tests performed to optimise the PCA algorithm. We then apply the optimum PCA algorithm to constrain the presence of water and clouds in the data via model comparison with a likelihood ratio test.
% This is followed by the model comparison performed to constrain the presence of water and clouds in the data, for which we used the optimum PCA algorithm variation as found here.

% ++++++++++++++++++++++++++++++++++++++++++++++++++++++++
% CC map
\begin{figure*}
\centering
\includegraphics[width=\textwidth]{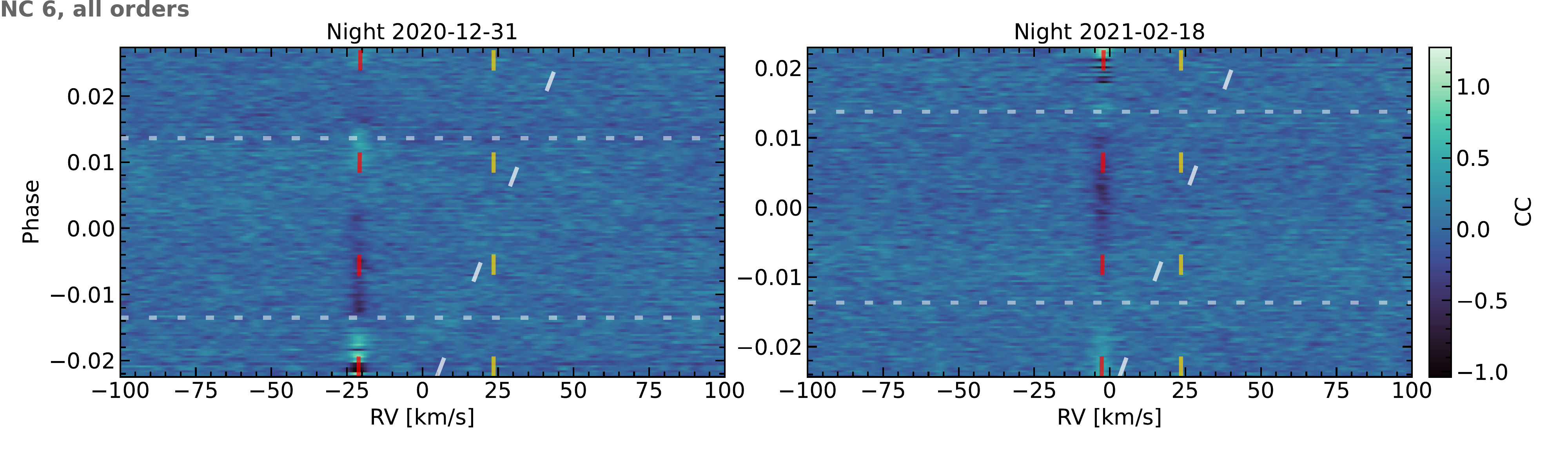}\\
\includegraphics[width=\textwidth]{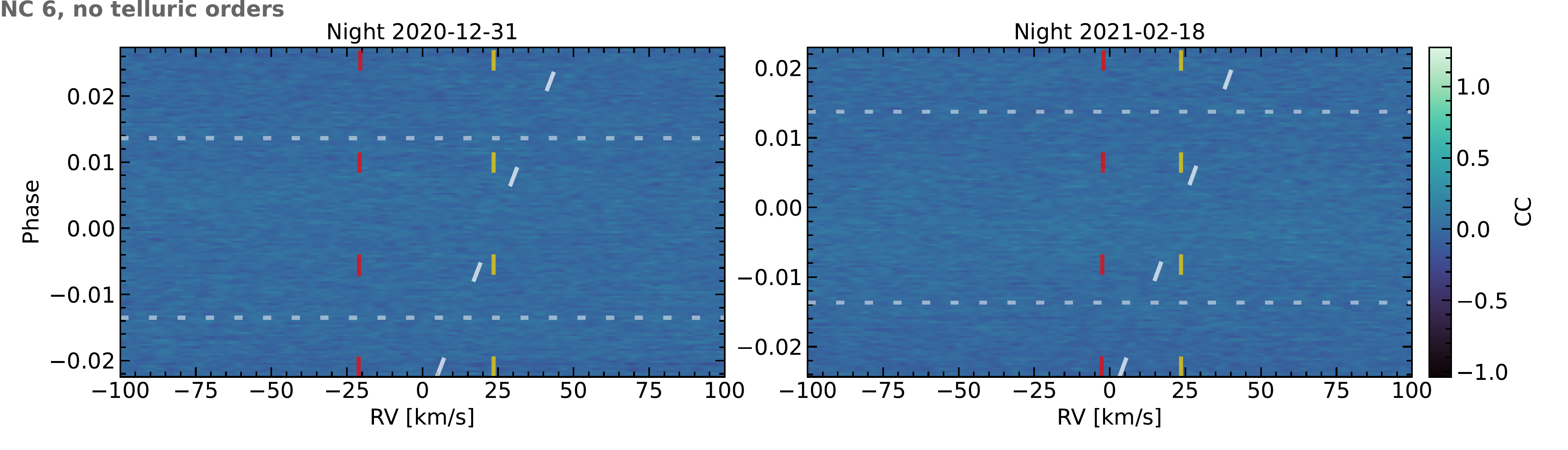}\\
\includegraphics[width=\textwidth]{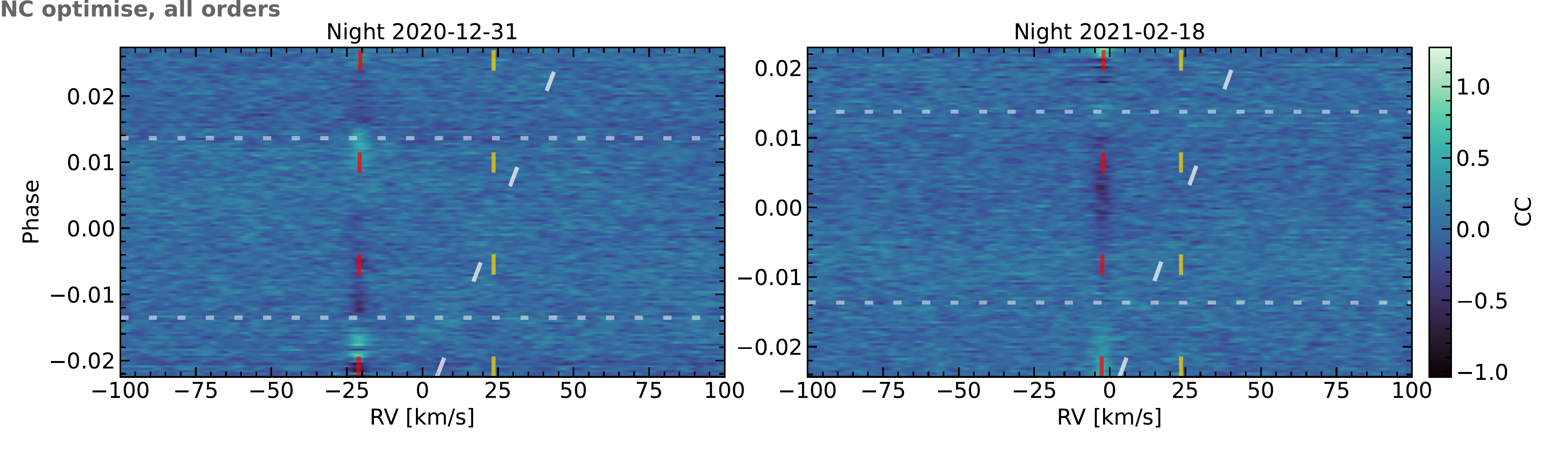}\\
\includegraphics[width=\textwidth]{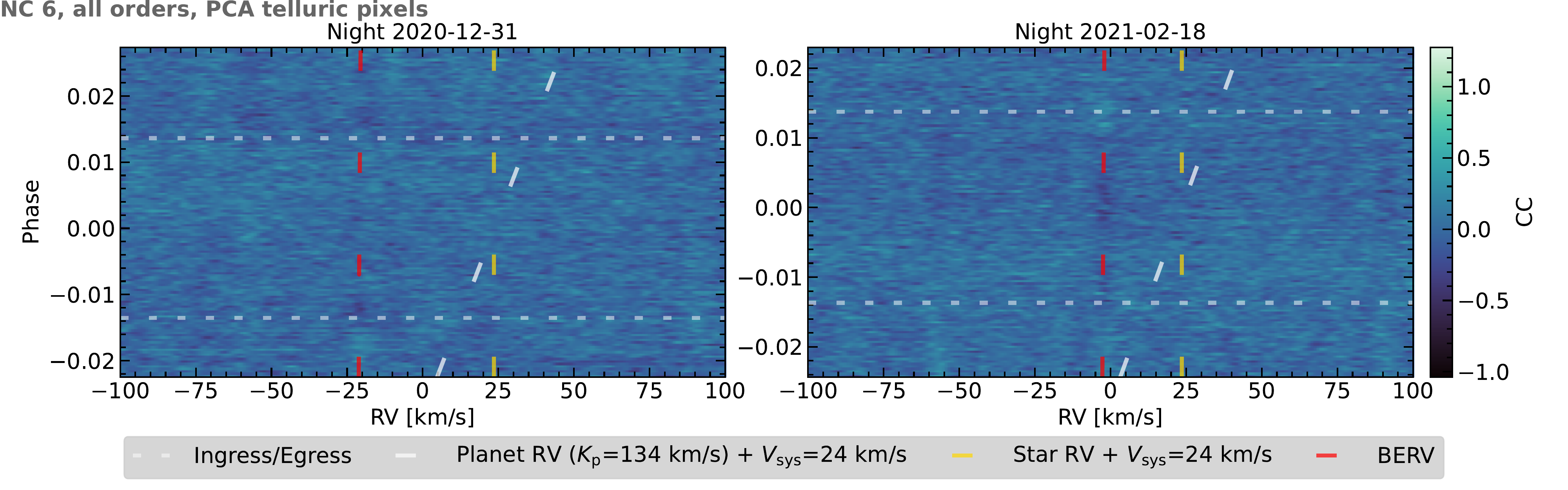}
\caption{Comparison of telluric removal algorithms (Section \ref{sec:pca} and sub-sections).
Top row: CC of each observation as a function of the observation phase. Left and right panels correspond to the two transits observed. The white, yellow, and red dashed lines correspond to the planetary, stellar, and barycentric Earth RVs, respectively. The short-dash white lines indicate the transit ingress and egress.
The CCs shown correspond to the coadding of the CCs of all the slices considered.
The CCs have been computed with the $\log _{10}(\mathrm{H_2O})=-3$ and $\log _{10}(P_\mathrm{cloud}\mathrm{/bar})=0$ model using the fast/unprocessed-model approach, and in the PCA processing we removed 6 components for all slices.
Second row: Same as top, but in this case only the CCs of slices 84-95, 104-107, 124-127, 142-145 (slices with no or small telluric contamination) have been coadded.
Third row: Same as top, but in this case the number of components removed per slice has been optimised.
Bottom row: Same as top, but in this case only the pixels affected by tellurics have been used in the PCA.
}
\label{fig:ccfmap_nc6}
\end{figure*}
% ++++++++++++++++++++++++++++++++++++++++++++++++++++++++

% ++++++++++++++++++++++++++++++++++++++++++++++++++++++++
% CC cut
\begin{figure*}
\centering
\includegraphics[width=\textwidth]{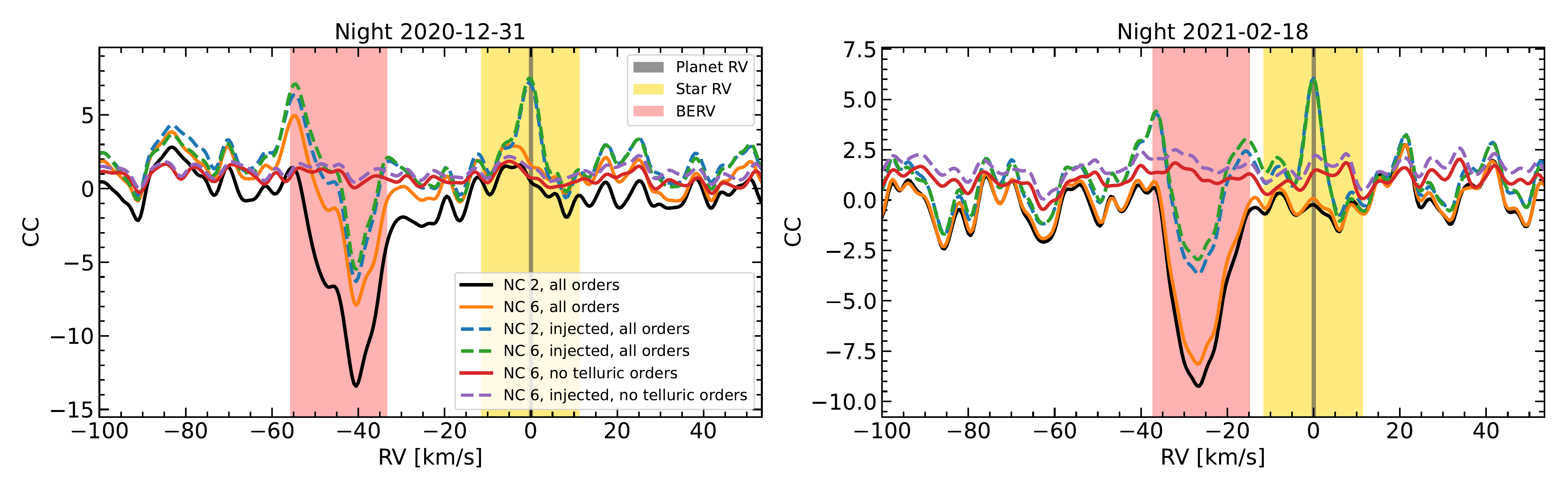}\\
\includegraphics[width=\textwidth]{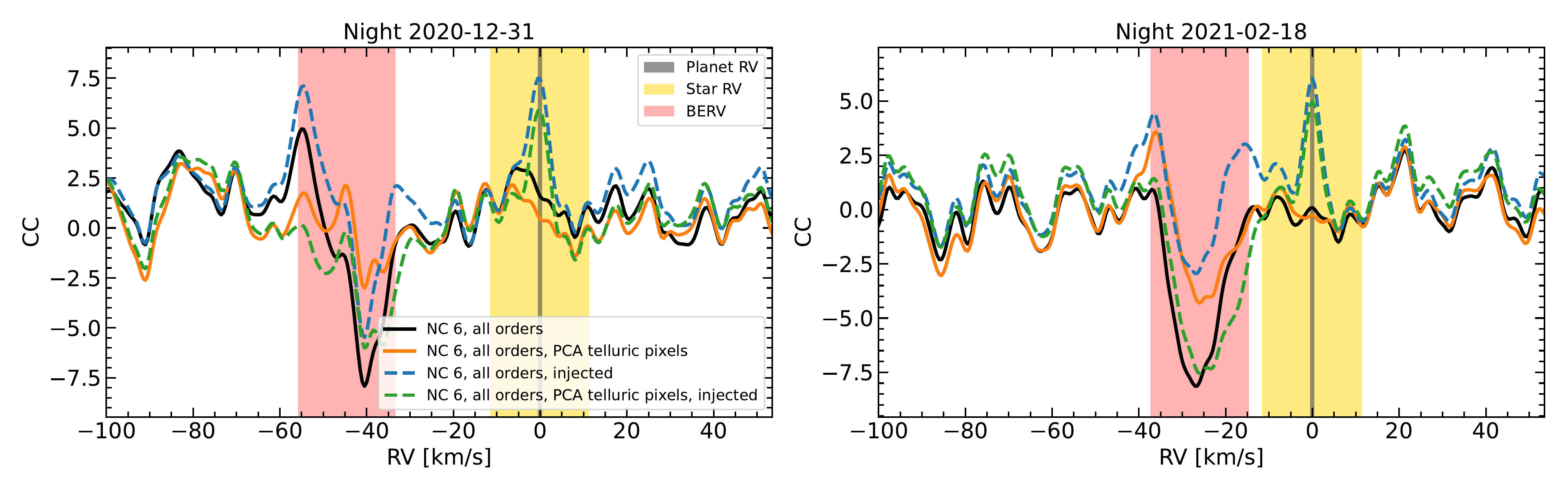}
\caption{CC functions from a selected sub-set of tests performed to optimise the telluric removal via PCA. The CC shown are the result of coadding the CC of each in-transit observation in the planet rest frame, at the expected \Kp and \Vsys.
Shaded grey, yellow, and red areas correspond to the the expected planetary, stellar, and barycentric Earth RVs, respectively.
Left and right correspond to the first and second night of observations, and top and bottom correspond to different sets of tests (see the legend).
}
\label{fig:cccut}
\end{figure*}
% ++++++++++++++++++++++++++++++++++++++++++++++++++++++++

% ++++++++++++++++++++++++++++++++++++++++++++++++++++++++
% Kp-Vsys map
\begin{figure*}
\centering
\includegraphics[width=\textwidth]{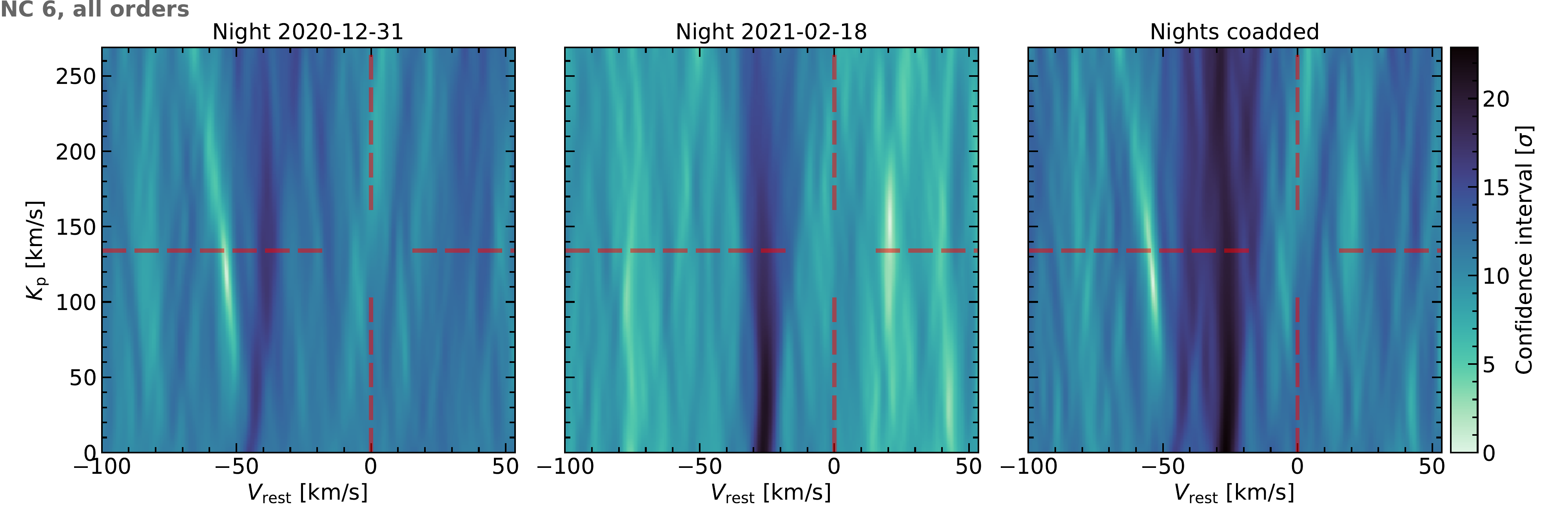}\\
\includegraphics[width=\textwidth]{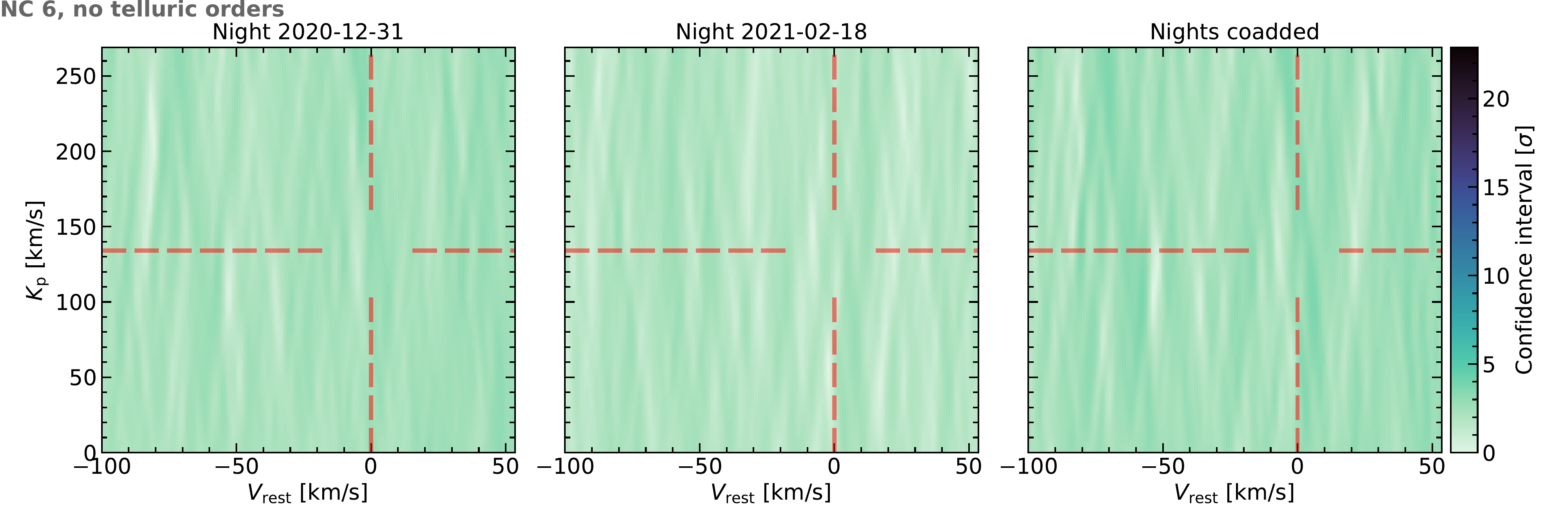}\\
\includegraphics[width=\textwidth]{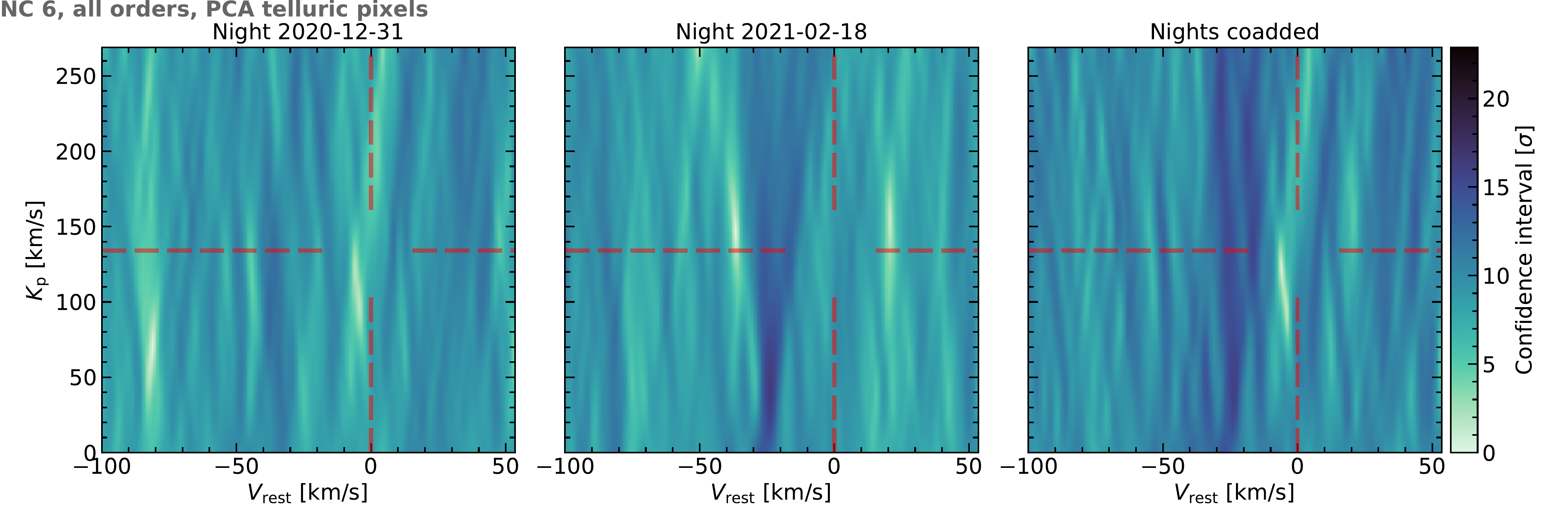}
\caption{
Top row: \KpVrest confidence interval maps obtained when removing 6 components from the PCA for all slices, using the $\log _{10}(\mathrm{H_2O})=-3$ and $\log _{10}(P_\mathrm{cloud}\mathrm{/bar})=0$ model based on the POKAZATEL line list to compute the CC, and coadding all the slices (this corresponds to coadding the in-transit CC shown in the top panel of Figure \ref{fig:ccfmap_nc6}). CCs computed using the fast/unprocessed-model approach.
Red dashed lines indicate the expected $K_\mathrm{p}$ and $V_\mathrm{rest}$.
Left to right are the maps for the first, second, and coadded nights.
Middle row: Same as top, but in this case, only the CCs of slices 84-95, 104-107, 124-127, 142-145 (slices with no or small telluric contamination) have been coadded (i.e. corresponds to coadding the in-transit CC shown in the above Figure \ref{fig:ccfmap_nc6}, second row).
% Third row: Same as top, but in this case the number of components removed per order has been optimised i.e. corresponds to coadding the in-transit CC shown in the above Figure \ref{fig:ccfmap_nc6}, third row).
Bottom row: Same as top, but in this case only the pixels affected by tellurics have been used in the PCA (i.e. corresponds to coadding the in-transit CC shown in the above Figure \ref{fig:ccfmap_nc6}, bottom row).
}
\label{fig:kpvsys_nc6}
\end{figure*}
% ++++++++++++++++++++++++++++++++++++++++++++++++++++++++

% ++++++++++++++++++++++++++++++++++++++++++++++++++++++++
% Kp-Vsys map, injected
\begin{figure*}
\centering
\includegraphics[width=\textwidth]{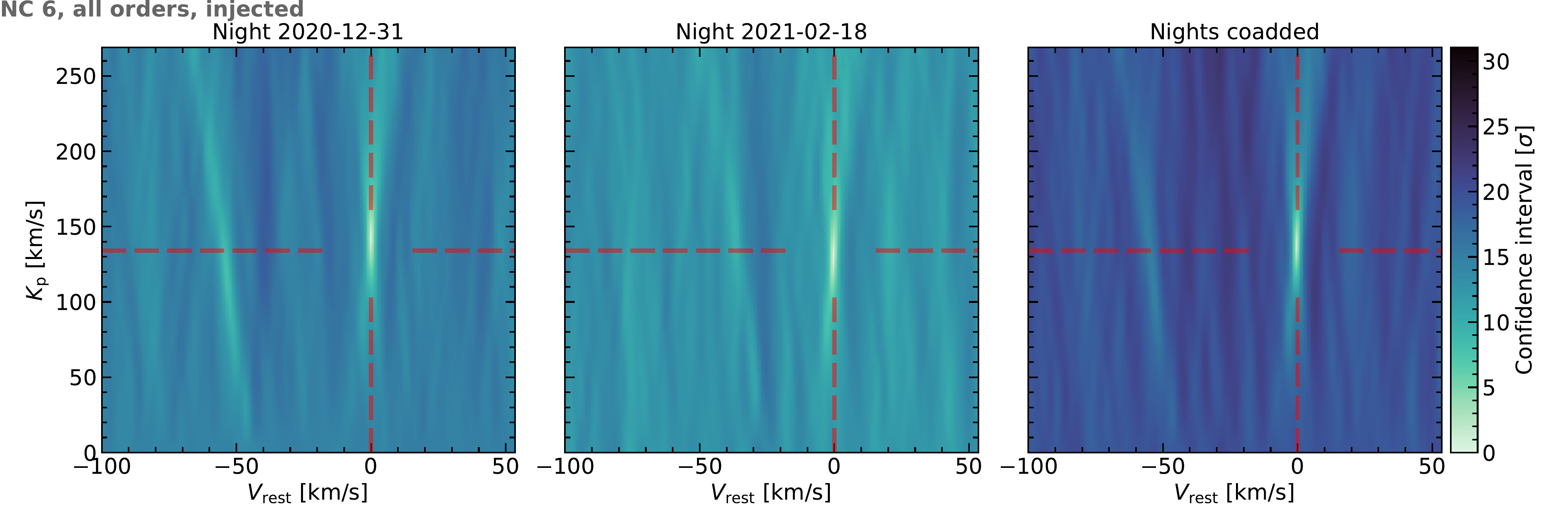}\\
\includegraphics[width=\textwidth]{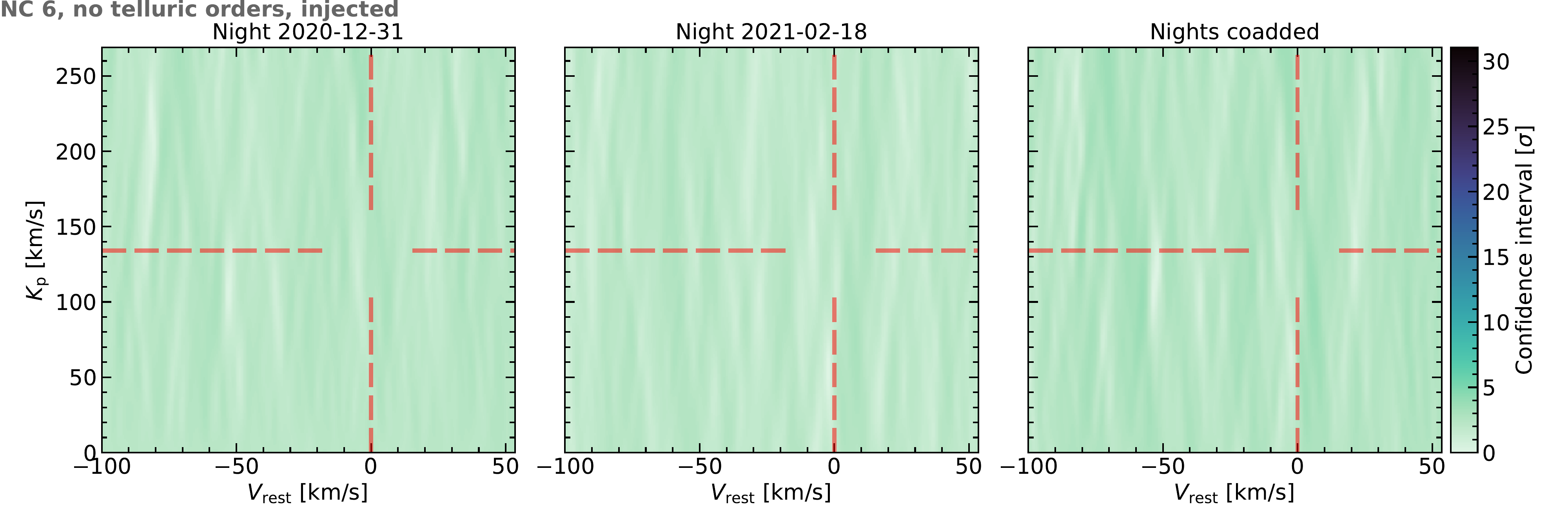}\\
\includegraphics[width=\textwidth]{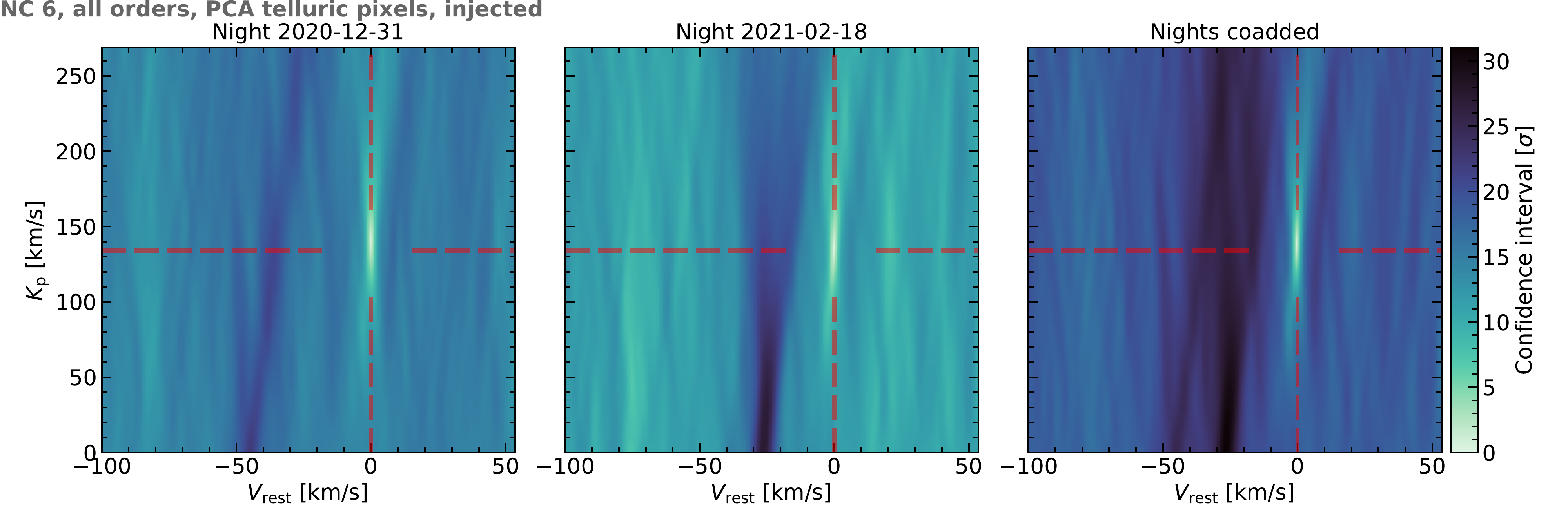}\\
\caption{Same as Figure \ref{fig:ccfmap_nc6}, but in this case, the planetary model used in the CC has been initially injected in the observations with no scaling.
}
\label{fig:kpvsys_nc6_inj}
\end{figure*}
% ++++++++++++++++++++++++++++++++++++++++++++++++++++++++

\subsection{PCA optimisation}

As explained in the methodology section (\ref{sec:pca}), we performed several tests with the goal of optimising the performance of the PCA to minimise the presence of tellurics in the CC and \logL functions. Here, we detail the results obtained.
Unless explicitly stated, all figures in this section display CC functions and \KpVrest maps obtained using the fast/unprocessed-model CC approach (Appendix \ref{sec:ccfast}).
This is because we wanted to directly study the shape of the CC functions, and in particular, the presence of telluric residuals, which is not possible with the slow/processed-model CC approach (Appendix \ref{sec:ccslow}).
Moreover, to cover the same \KpVrest parameter space, the slow/processed-model method takes significantly longer computational time than the fast/unprocessed-model one, which in practice limits the \KpVrest values that we can sample, as well as the number of tests we can do.
Therefore, to perform our initial tests, we decided to follow the first approach. This allowed us to test the optimal parameters for the PCA and identify any tentative planetary signals.
% However we note that, for the model comparison (Section \ref{sec:modelcomparison}), we followed the second CC approach with a reduced \KpVsys range around the expected planetary signal, covering any tentative signals we observed in our initial tests. We also performed the same grid analysis with the first approach to check for any differences.

\subsubsection{Fixed number of PCA components}\label{sec:results_ncfix}
% \subsubsection{Number of PCA components and coadded orders}

Applying the PCA algorithm removing a fixed number of components for all the slices results in a strong telluric feature in the CC functions. We show this in the top panels of Figure \ref{fig:ccfmap_nc6} and Figure \ref{fig:cccut} (black and orange lines), where we indicate the position of the telluric residuals in red.
Figure \ref{fig:ccfmap_nc6} shows the CC functions obtained for all the observations as a function of the orbital phase for the two nights (columns), for different tests performed (rows). Figure \ref{fig:cccut} shows the in-transit CC functions coadded in planet rest frame for the two nights (columns) and different tests (different lines in both rows).
We know that the observed feature is due to telluric contamination because it appears at the expected BERV and spans the entire sequence of observations (i.e. it is not phase dependent and is present in both in-transit and out-of-transit observations).
In the CC functions, we see that the signal evolves in time from correlated (maxima) to anti-correlated (minima), as a result of the positive and negative residuals in the processed spectra. These residuals, in turn, come from the change in airmass and the changes in the atmospheric integrated water vapour column that changes during the observations, which increase and decrease the amount of water vapour above or below the overall trend. 
% Hence, the residuals seen in the CC functions vary in sync with the water vapour changes and airmass. 
The telluric residuals do not perfectly correlate with changes in airmass and water vapour because we have applied the PCA and removed the first components prior to computing the CC functions. That is, the first PCA components removed contain part or most of the airmass and water vapour variations, and hence, the correlation is broken.
We do not show them here, but if we calculate the same maps using the \logL function (Equation \ref{eq:cc2logL}) rather than CC (Equation \ref{eq:r2cc}), we see the same residuals.
%the same residuals are seen in the \logL functions.
% - Could CC with Earth telluric model to enhance that.
The telluric feature is also clearly seen in the form of maxima and minima in the \KpVrest maps produced after coadding the in-transit \logL functions in planet-rest-frame, see e.g. top panels in Figure \ref{fig:kpvsys_nc6}, where we plot the confidence intervals obtained for each night and for both nights coadded (columns).

As expected, the telluric signal decreases as we increase the number of components removed during the PCA, see top panels in Figure \ref{fig:cccut} for examples removing 2 and 6 components (black and orange lines), but it is never completely removed. 
We notice that the removal seems to work better in the first night than in the second one, i.e., when the integrated water vapour is higher.
We tested removing between 2 to 15 components on all the slices, but found no significant improvement, i.e., the telluric signal did not decrease further, after removing more than $\sim 6$ components.
We qualitatively explain the inability of PCA to de-trend telluric lines as follows. In optical observations where telluric lines are not prominent, their contribution to the total variance within one slice is also negligible. Since the SVD algorithm `ranks' components based on their contribution to the variance, telluric residuals might potentially be absent in the first 15 components, which would instead be dominated by throughput and continuum variations. 
The residual level of correlation is similar to that expected from standard telluric removal algorithms e.g. direct modelling of the telluric spectrum, unless these residuals are heavily masked prior to correlation. To improve the correction, we revised the SVD algorithm as explained in Section \ref{sec:selected_feeding}.
%We address this aspect by altering the SVD algorithm as in Section \ref{sec:selected_feeding}.

% --- Injected
\paragraph*{Injection tests}
We also tested the behaviour of the PCA algorithm when injecting a planetary model (water abundance $\log _{10}(\mathrm{H_2O})=-3$, in VMR, and cloud deck pressure $\log _{10}(P_\mathrm{cloud}\mathrm{/bar})=0$ based on the POKAZATEL line list) in the data.
The \KpVsys maps (Figure \ref{fig:kpvsys_nc6_inj}, top) show that an injected planetary signal with a scaling of $\times1$ (i.e., original strength) is recovered with high significance, despite the presence of tellurics in the data.
The injected signal is clear in each night individually, with a higher confidence in the first night that increases when combining both nights.
From the CC maps (Figure \ref{fig:ccfmap_nc6}), we see that in both nights, the expected planetary RV and the BERV do not overlap, which might help in obtaining a significant detection.
Even when only removing 2 PCA components, the injected planetary signal is clearly detected in each night in the form of a peak in the CC and \logL functions, see top panels in Figure \ref{fig:cccut}, where we compare removing 2 and 6 components (blue and green dashed lines, respectively).
From these same tests we also see that the injected planetary signal is not affected by increasing the number of components removed. This indicates that the PCA components are not selecting the injected planetary signal, which is the behaviour we expect. 
We also note from Figure \ref{fig:cccut} that the telluric residuals in the CC are slightly different if the model has been injected in the data (black, orange lines) or not (blue, green dashed lines), which indicates that part of the injected planetary signal could impact the PCA, despite the fact that its significance does not decrease.

% --- Removing telluric orders
\paragraph*{Neglecting orders affected by tellurics}
The orders where the telluric effect is stronger are those for which we see strong telluric absorption lines. These orders are also those where the planetary water shows the strongest absorption lines (see e.g. Figure \ref{fig:ord_tpl_model}).
Coadding the CC (or \logL) functions discarding these telluric-affected order slices (i.e., using only slices 84-95, 104-107, 124-127, 142-145) results in a decrease in the telluric signal, see the CC functions in the second row of Figure \ref{fig:ccfmap_nc6} and the top panels in Figure \ref{fig:cccut}, red line, which show no significant feature at the telluric position in RV space.
The telluric residuals also disappear in the \KpVrest maps, see middle panels in Figure \ref{fig:kpvsys_nc6}.
% All points in these maps are within less than $\sim4~\sigma$ from each other, which means that none of the residuals are significant.
We note here that in these maps, all data points are within $\sim4~\sigma$ (or less) of one another. This means that none of the data points, i.e., none of the \KpVrest pairs, is more significant than any other. In other words, the points with the highest likelihood in maps without the telluric orders maps (i.e. confidence interval close to 0) are not significant.
% In comparison, for the other cases shown also in Figure \ref{fig:kpvsys_nc6} (top and bottom rows), the data points corresponding to telluric residuals are over $\sim15~\sigma$ away from the points with highest likelihood (see the colour bars). 

If we now look at the cases where we have injected a water model, we note that the planetary signal that is clearly detected using all the orders also disappears. We see this in the top panels of Figure \ref{fig:cccut}, purple dashed line, where the clear signal at the injected RVs is no longer there, and the CC looks as flat as in the case where we have not injected a planet, as well as in the \KpVrest in the second row of Figure \ref{fig:kpvsys_nc6_inj}. As happens in the case without any signal injected, now all data points in the \KpVrest maps are also within $\sim4~\sigma$ of one another, meaning that no data point is significant with respect to the others.
The fact that the injected planetary signal is not seen here is not surprising. Despite the fact that the exoplanet temperature is significantly higher than the Earth's temperature, the main \water features are similar, and thus removing orders containing telluric \water also removes exoplanetary \water.

%%%%%%%%%%%%%%%%%%%%%%%%%%%%%%%%%%%%%%%%%%%%%%%%%%

\subsubsection{Optimisation of the number of PCA components per slice}
% \textcolor{red}{[What happens with each order? For which ones do we reach the maximum and there is still a signal?}

We also optimised the number of components to be removed per slice using the method described in Section \ref{sec:ccopt}.
The slices that are optimised, i.e., those that result in an increased number of components removed with respect to the initial, are in general those that contain strong telluric absorption lines: slices 80 - 83, 98, 99, 108 - 111, 116 - 121, 134 - 139, 147 - 155, 158 - 161, 168, 169, see Figure \ref{fig:ord_tpl_model} above. 

There is a relatively strong telluric band covering slices 128 - 131, and the strongest band of saturated O$_2$ lines in slices 162 - 165, for which the number of components are not optimised. In the case of the saturated band in slices 162 - 165, it is possible that the telluric lines are simply too strong for the PCA to be able to remove its effect, however this argument does not explain why the weaker band in slices 128-131 is not being properly removed. Further analysis is needed to understand these results and the behaviour of the PCA algorithm.

In most of the order slices that contain tellurics, both slices have an increased number of tellurics removed, but the final number of components is not always the same for both slices of the same order. This is not necesarrily expected, and suggests that the PCA is selecting additional correlated noise different in both slices, rather than purely telluric signals, which should be the same in both slices. Again, further analysis of the PCA behaviour is needed to understand this difference.

For the two nights, most of the slices mentioned above are optimised. However, the final number of components also differs between nights for the same slices, which is expected since the tellurics behave differently in the two nights.
% Several \textbf{slices} reach the maximum number of components considered (in this case, 15).

Figure \ref{fig:ccfmap_nc6}, third row, shows the CC functions obtained when applying this optimisation. We notice that, despite removing a significantly higher number of components for the telluric-affected orders, this results is a CC map very similar to the one we obtain by removing a lower, fixed number of components.
The \KpVrest map is also similar to this case, hence we have not included it here.
This indicates that, despite still having strong telluric residuals in the CC, removing a higher number of components does not result in a significant telluric removal.

%%%%%%%%%%%%%%%%%%%%%%%%%%%%%%%%%%%%%%%%%%%%%%%%%%

\subsubsection{Selectively feeding telluric lines into the PCA}\label{sec:selected_feeding}

As explained in Section \ref{sec:tellpca}, we modified the PCA algorithm to focus on the pixels affected by telluric lines, rather than using the full spectral order. We show the CC function and \KpVrest maps that we obtain in the bottom panels of Figures \ref{fig:ccfmap_nc6} and \ref{fig:kpvsys_nc6}.
In the CC function maps, we see some telluric residuals at the beginning of the first night, including part of the transit, as well as some faint residuals during the transit of the second night. This translates into a very faint signal in the \KpVrest map at the position where we expect the telluric residuals to be for the first night, and in a stronger residual for the second one.
Compared to the results obtained using the full spectral order (top panels in Figures \ref{fig:ccfmap_nc6} and \ref{fig:kpvsys_nc6}), in this case, the telluric contamination is significantly removed in both the CC functions and \KpVrest maps.
This means that the PCA components removed are more effective in tracing the telluric variability if we only use the regions of the spectrum affected by tellurics rather than the whole wavelength range.
This is again expected, because in the (sub-)matrix containing only telluric lines the latter will have a more noticeable contribution to the variance, and therefore will be ranked higher by the SVD algorithm.

% --- PCA NC/order optimisation
In addition to using the telluric-affected pixels in PCA, we also performed the optimisation of the number of components to be removed per slice, as done above.
In this case, since the initial PCA components are already removing most of the telluric signal, the optimisation algorithm did not detect a signal strong enough to continue removing components. Hence, for almost all the slices, the algorithm stops after the initial number of components considered has been removed. 
%Hence, the CC and \KpVrest maps applying the optimisation or not look very similar (and are not shown here).
This means that the CC and \KpVrest maps look very similar if we apply the optimisation and if we do not, and are not shown here.
% Few orders have more than 6 components removed.
% Of the order pairs, only 1 has a decrease in the number of components removed, which indicates that the signal that is picked up as telluric is not strong since it is not seen in both orders.

% --- Injected
\paragraph*{Injection tests}
With the new SVD algorithm, injected planetary signals are still recovered at high significance, see bottom panels of Figure \ref{fig:kpvsys_nc6_inj}.
As happened when using the full spectral range (Section \ref{sec:results_ncfix}), the telluric residuals look different if the planetary signal has been injected in the data or not; see bottom panels of Figure \ref{fig:cccut}, orange and green lines, and bottom panels in Figures \ref{fig:kpvsys_nc6} (non-injected) and \ref{fig:kpvsys_nc6_inj} (injected case). The telluric residuals are stronger if the planetary signal has been injected.
Similarly, if we now compare the case where only the telluric-affected regions are used in the PCA with the initial case where the full spectral range of the order is used (both with an injected planetary signal), the telluric residuals are different, see again Figure \ref{fig:cccut}, bottom, and top and bottom panels in Figure \ref{fig:kpvsys_nc6_inj}.
In general, for both nights, the tellurics are more significant if only the telluric-affected pixels are used in the PCA (bottom panels of Figure \ref{fig:kpvsys_nc6_inj}) compared to the whole spectral order (top panels of Figure \ref{fig:kpvsys_nc6_inj}), which is the opposite as what happens when no planetary signal is injected.
As mentioned before, this indicates that the injection of a planetary model affects the telluric identification in the PCA algorithm, but this does not seem to affect the planetary signal itself, as it is clearly detectable in both cases.

% --- Signal?
\subsubsection{A tentative H$_2$O signal from WASP-166~b?}

The \KpVrest map of the first night obtained with the analysis in Section \ref{sec:selected_feeding} above (i.e. with the modified PCA algorithm) shows a correlated signal close to the expected planetary position, about 5~\kms blue-shifted from the expected \Vrest and extending about $-30$~\kms from the expected \Kp (bottom left panel in Figure \ref{fig:kpvsys_nc6}).
The signal is slightly significant with respect to its neighbouring points.
While this is outside of the uncertainties of \Vrest (or \Vsys) and \Kp (see Table \ref{tab:sysparams}), unaccounted atmospheric circulation at the \kms level has been shown to potentially alter \Vrest and \Kp measurements.
The second night shows a similar structure but less prominent and not significant. This could be affected by the fact that the tellurics are less removed in the second night than in the first one, and hence, a possible planetary signal might be hidden in the telluric residuals. In addition, in RV space, the tellurics are closer to the expected \Vsys in the second night compared to the first one.
Despite this difference, the signal is still present when coadding both nights. It is also more significant with respect to its neighbouring points than in the first night alone. We will further discuss this candidate signal in Section \ref{sec:modelcomparison_results}.

%%%%%%%%%%%%%%%%%%%%%%%%%%%%%%%%%%%%%%%%%%%%%%%%%%

% \subsubsection{Summary}

% In summary, using the modified PCA algorithm in which we use only the spectral regions affected by tellurics results in less significant telluric residuals than any of the other tests performed.
% With the injection tests we see that, if we discard the orders that are significantly affected by telluric contamination, the injected planetary signal decreases in significance, while using all the orders results in a significant detection, even in the presence of strong tellurics in the processed data.

%%%%%%%%%%%%%%%%%%%%%%%%%%%%%%%%%%%%%%%%%%%%%%%%%%

\subsection{Model comparison}\label{sec:modelcomparison_results}

% ++++++++++++++++++++++++++++++++++++++++++++++++++++++++
\begin{figure*}
\centering
\includegraphics[width=\textwidth]{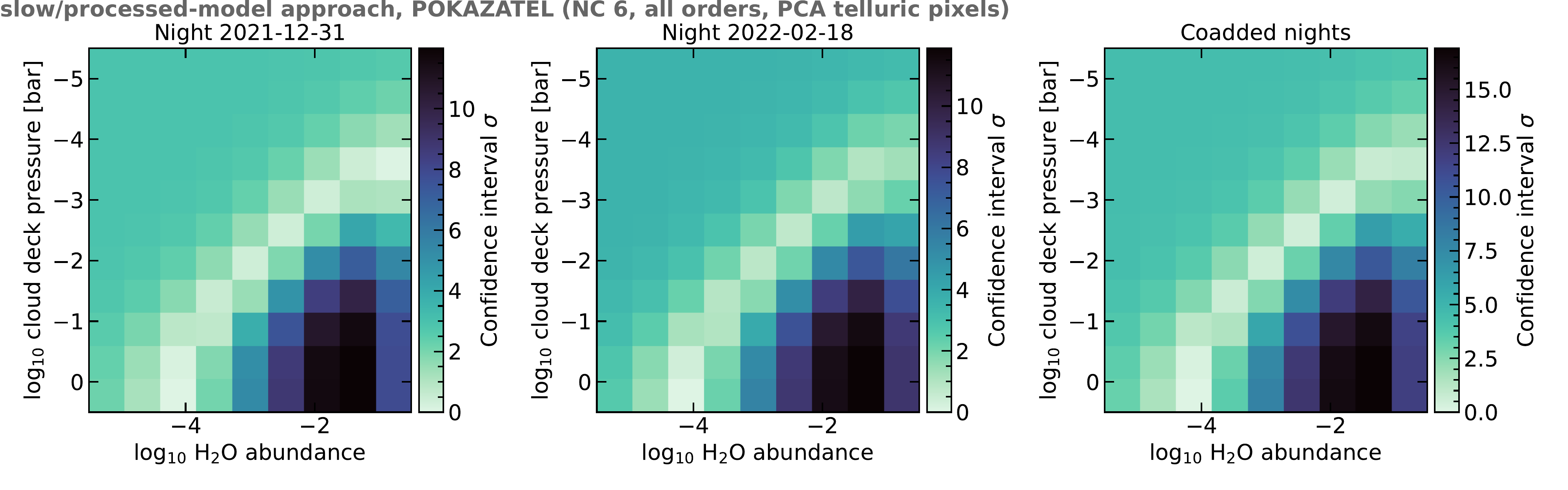}\\
\includegraphics[width=\textwidth]{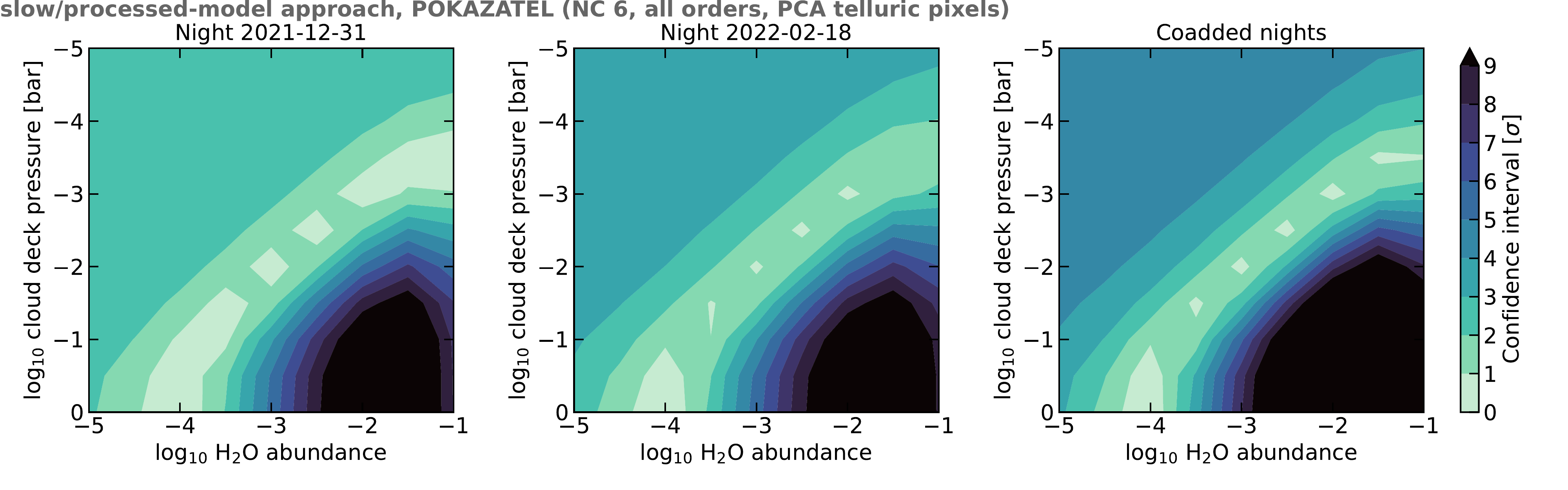}
\caption{Top: Confidence intervals (colour) for the POKAZATEL model grids of different water abundances ($\log_{10}(\mathrm{H_2O}) = -1$ to $-5$, in VMR, x-axis) and cloud deck pressures ($\log_{10}(P_\mathrm{cloud}/\mathrm{bar}) = 0$ to $-5$, y-axis) obtained with the slow/processed-model approach to compute the CCs. Left and middle panels are the results for each of the transits, and right, for both transits coadded.
Bottom: Same as top, but here contour levels have been added to help visualise the confidence intervals.}
\label{fig:model_comparison}
\end{figure*}
% ++++++++++++++++++++++++++++++++++++++++++++++++++++++++

% ++++++++++++++++++++++++++++++++++++++++++++++++++++++++
\begin{figure*}
\centering
\includegraphics[width=\textwidth]{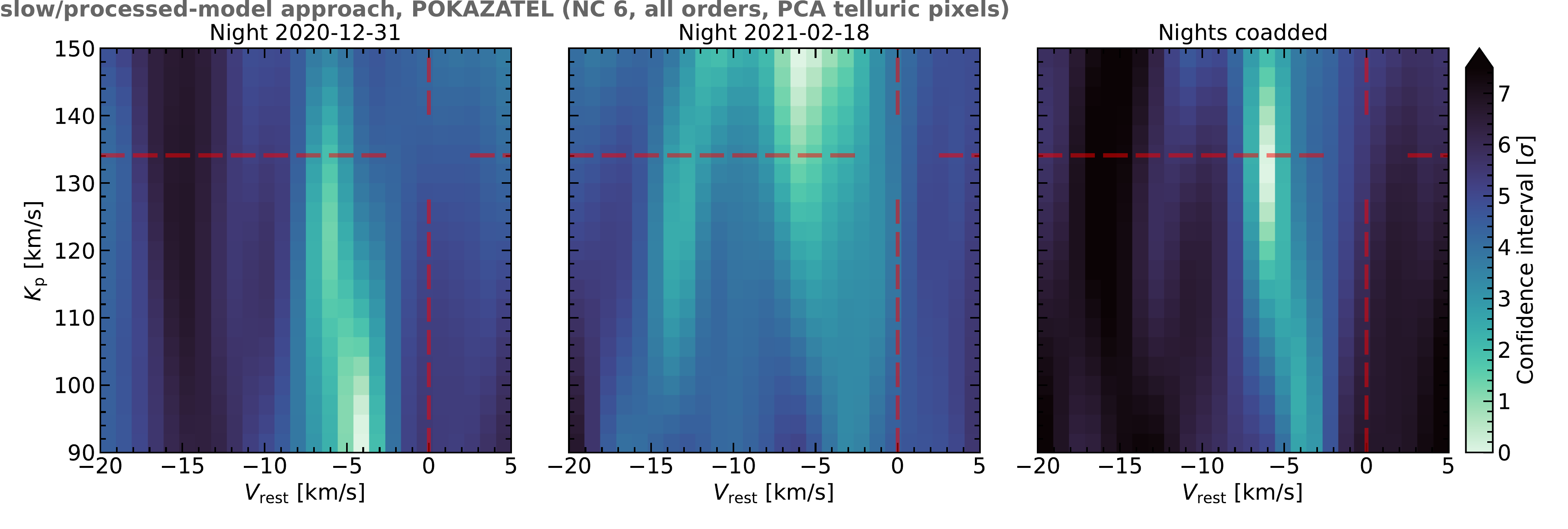}\\
\includegraphics[width=\textwidth]{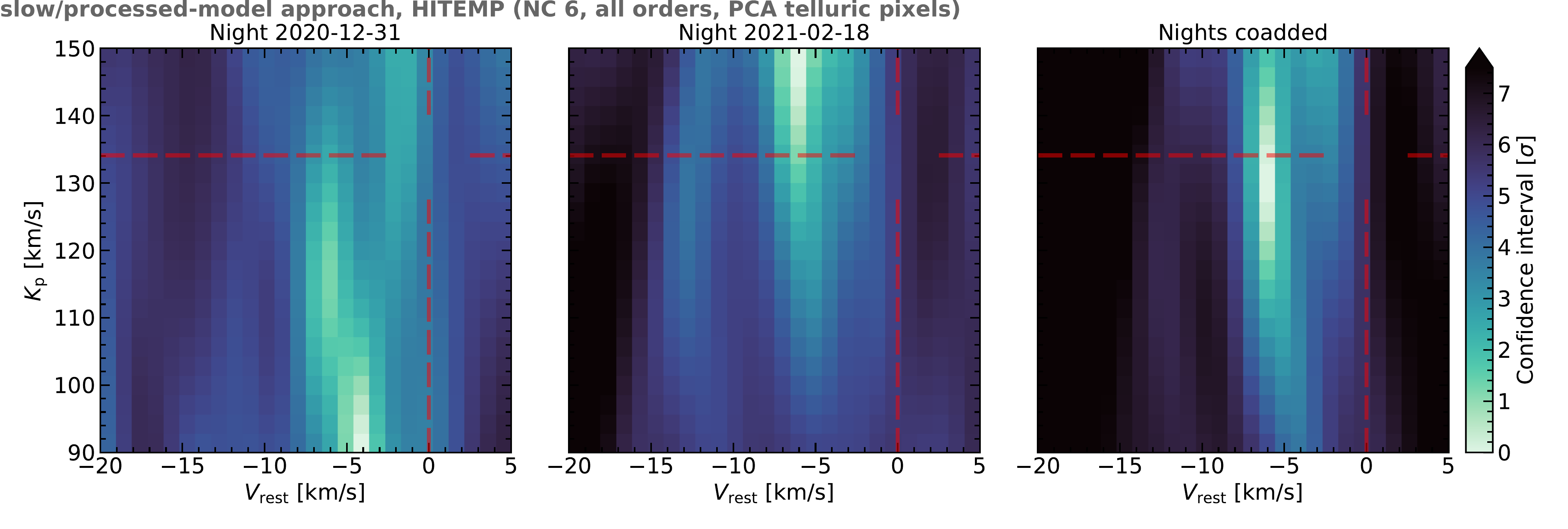}
\caption{
\KpVsys confidence interval maps obtained with the PCA algorithm that uses only telluric-affected regions, removing 6 PCA components, CCs computed with the slow/processed-model approach, and using the $\log _{10}(\mathrm{H_2O})=-4$ and $\log _{10}(P_\mathrm{cloud}\mathrm{/bar})=0$ model based on the POKAZATEL (top) and HITEMP (bottom) line list to compute the CC, and coadding all the order slices.
Red dashed lines indicate the expected $K_\mathrm{p}$ and $V_\mathrm{sys}$.
Left to right are the maps for the first, second, and coadded nights.
}
\label{fig:slow_fast_comparison}
\end{figure*}
% ++++++++++++++++++++++++++++++++++++++++++++++++++++++++

In the previous section, we see that the modified PCA algorithm in which we use only the spectral regions affected by tellurics in the SVD results in less significant telluric residuals than any of the other tests performed.
% The injection tests show that, if we discard the orders that are significantly affected by telluric contamination, the injected planetary signal decreases in significance, while using all the orders results in a significant detection, even in the presence of strong tellurics in the processed data.
Therefore, to perform the model comparison between different theoretical models (as explained in Section \ref{sec:modelcomparison}), we used the data processed using the modified SVD algorithm, since it minimises the telluric residuals.
Moreover, to be able to compare the different models, it is important to correctly reproduce the line strength of the data. We can only guarantee this if the model used in the CC has been processed by the same PCA as the data, which we do here by using the slow/processed-model CC approach.

% Model grid
\subsubsection{Grid search} \label{sec:results_gridsearch}

Figure \ref{fig:model_comparison} shows the confidence intervals obtained for the grid of 99 models tested. These results correspond to the models created with the POKAZATEL line list, but we obtain equivalent results for the HITEMP line lists.
In each night separately, and also when coadding both nights, models with a high content of water and a cloud deck at high pressure (\logwater$\lesssim -3$ and \logcloud$\gtrsim -2$, bottom-right quadrant of Figure \ref{fig:model_comparison}) are rejected with high confidence compared to the other models tested ($\gtrsim 6~\sigma$, up to $15~\sigma$ for some models).
Models with the lowest cloud deck pressures and lowest water abundances (upper-left quadrant of the plot), are also excluded but only with $\sim 4~\sigma$ confidence.

Overall, the preferred model is that with \logwater$=-4$ and \logcloud$=0$ (i.e. no cloud deck).
There is a preference for intermediate models with low water content and high cloud deck pressure, or higher water content and lower cloud deck pressure (models coloured in light-green in Figure \ref{fig:model_comparison}). These models are within a confidence interval of $\sim2~\sigma$ of the preferred model. This happens in all cases, i.e., for the two nights individually and both nights coadded.

The fact that the intermediate models are preferred over those with the lowest cloud deck pressure and lowest water content (upper-left quadrant of the plot) points towards a tentative detection of a water signal. If there was no planetary signal present, the preferred models will be those that have the shallowest absorption lines, i.e., those that are compatible with an almost flat spectrum (see models in Figure \ref{fig:models}). These are the models with the lowest water abundance and low-pressure clouds, i.e, models in the upper-left quadrant of Figure \ref{fig:model_comparison}), which are not preferred here. In other words, a non-detection would only exclude the bottom-right quadrant of the grid, but not the upper-left, as happens here. This is in qualitative agreement with the predictions of \citet{gandhi2020clouds}.

We note that at low cloud pressures, models with water abundance \logwater$\simeq-2$ are more strongly rejected than those with higher water abundances (\logwater$=-1$). 
This is due to the higher mean molecular weight of the atmosphere with \logwater$=-1$ compared to the one with \logwater$=-2$. As we increase the abundance of water, for \logwater$\gtrsim-2$, the mean molecular weight of water-rich atmospheres becomes higher than at lower water abundances. This higher mean molecular weight reduces the scale height, which results in a decrease in the strength of the water absorption features (see Figure \ref{fig:models_mu} in Appendix \ref{sec:models_mu}). Hence, due to the fact that for water abundances \logwater$\simeq-2$ the absorption features are stronger than for \logwater$=-1$, models with \logwater$\simeq-2$ are more strongly rejected \citep[also see e.g.][]{gandhi2020clouds}.

Our confidence interval analysis is relative to the `best' model (the one with a highest likelihood). That is, the best model has by default a $\sigma$ of 0 and the other models have then $\sigma$ values relative to the best one. In general, it is not clear how to assess the absolute significance of the best model. Even commonly-used signal-to-noise approaches do not assess how good a model fits to  the data in an absolute sense.
To try to obtain a `baseline' likelihood and assess the absolute significance of our models, we have performed an extra test with a `flat' model, i.e., a model with flux equal to 1 at all wavelengths. We have used this flat model to compute the CC and \logL functions with the data processed with the best PCA algorithm using the slow/processed-model method (including model processing), as done with the model grid above. We have then compared the \logL obtained with this model with that of the `best' model according to our grid analysis (\logwater$=-4$ and \logcloud$=0$) in the same way as we compared the different models in the grid above. That is, we performed a likelihood ratio and computed how many $\sigma$s away the models are from each other. For the first night, the flat model is rejected with 3.9~$\sigma$ compared with the best one, for the second night, 4.2~$\sigma$, and for both nights coadded, 4.9~$\sigma$.
This tells us that in the data, there is a signal (regardless of its origin, planetary or telluric) that is $\sim5~\sigma$ above a flat model.
This test is similar to comparing the best model with models close to a flat line (i.e. those in the top left corner of our grid). Indeed, the difference in sigma obtained between the best model and those in the top left corner is similar to that obtained with respect to the flat model.

% Single model example
\subsubsection{\KpVsys maps of the preferred model}

In Figure \ref{fig:slow_fast_comparison} we show a comparison of the \KpVrest maps obtained with the model favoured by our grid search in the previous section (model with \logwater$=-4$ and \logcloud$=0$), obtained with each of the two line lists considered, POKAZATEL (top) and HITEMP (bottom).
As mentioned above, here we used the slow/processed-model CC approach, including model processing, as opposed to the results we show in the bottom panels of Figure \ref{fig:kpvsys_nc6}, in which we use the fast/unprocessed-model CC approach. Note also the smaller \Kp and \Vrest ranges explored here with the slow/processed-model approach, which are around a blue-shifted signal close to the expected planet position.

This blue-shifted signal close to the expected planet position, but with lower \Kp value than expected, was already seen in the initial tests with the fast/unprocessed-model CC approach (Figure \ref{fig:kpvsys_nc6}, bottom panels) and is still present in the first night.
For the second night, this signal was less significant than in the first night in the initial tests. Now, with the slow/processed-model CC approach including the model processing, a signal also blue-shifted $\sim5~\kms$ appears, but it is shifted towards higher \Kp values. 
This difference between the fast/unprocessed-model and slow/processed-model approach is something expected. With the fast/unprocessed-model method, we process the data through the PCA and then directly cross-correlate it with a model. However, in the slow/processed-model approach, we additionally process the model through the same PCA as the data before computing the cross-correlation. By doing this extra processing of the model, we modify the model in the same way as the data has been modified by the PCA. 
% Therefore, the slower approach should result in a better fit than the fast one.
The fact that this signal is clearer in the second night using the slow/processed-model approach highlights the importance of model processing. By altering the model in the same way the PCA has altered the data, the CC should result in a better match between data and model, which is what we see here.

Coadding both nights results in a $\sim5~\kms$ blue-shifted signal at the expected \Kp, which may be the result of the original signals in both nights being displaced in \Kp in opposite directions. This blue-shifted signal is favoured with respect to the neighbouring points with $\sim 7~\sigma$.
For both line lists, the results obtained are equivalent, i.e., both nights show a blue-shifted signal displaced from the expected \Kp, with very similar \KpVrest maps, and the signal is still present at high significance when both nights are coadded.
Despite being at different \Kp, the individual night signals are within 1 and 2~$\sigma$ (for the second and first night, respectively), of the coadded nights signal. Therefore, the signal observed when coadding both nights is not rejected by the results obtained for each night individually.
We note that \Kp is not strongly constrained in the transit observations that we are considering here, because they only span a small part of the total Keplerian orbit. Hence, it is hard to obtain good constraints in \Kp, which could be the cause of the discrepant \Kp values observed in these maps.

The $-5~\kms$ shift in rest-frame velocity is outside of the uncertainties on the measured \Vrest (or \Vsys), which has been obtained from the same observations as we use here (see Table \ref{tab:sysparams}).
The observed blue-shift could potentially be due to the presence of winds in the planetary atmosphere.
The planetary Na lines detected by \citet{seidel2022wasp166} in the same observations show a significant broadening with respect to the instrument profile, of 9.37$\pm$0.95~\kms, which suggest that the Na is moving at high velocities, similarly to what we might be seeing here with \water. 
% If the atmosphere is cloud free, i.e., if the Na is probing down to the surface, the velocity would not be enough for the Na to be escaping (the escape velocity at the surface is of $\sim$24~kms).

The telluric residuals for the first and second nights are at $\sim -$50~\kms and $\sim -$25~\kms of this signal, respectively. Therefore, we do not expect the pixels neighbouring the blue-shifted signal to be significantly affected by tellurics.

%%%%%%%%%%%%%%%%%%%%%%%%%%%%%%%%%%%%%%%%%%%%%%%%%%
%%%%%%%%%%%%%%%%%%%%%%%%%%%%%%%%%%%%%%%%%%%%%%%%%%

\section{Summary and concluding remarks}\label{sec:conclusion}

In this work, we have analysed two transits of the inflated super-Neptune WASP-166~b observed with the optical, high-resolution spectrograph ESPRESSO. Using the high-resolution cross-correlation technique, we study the presence of water vapour and clouds in the atmosphere of the planet.

% PCA
To correct for the presence of telluric signals which may interfere with a potential planetary signal, we start by applying a PCA algorithm on the observed spectra. We noticed that a standard PCA algorithm results in very strong telluric residuals in the CC and \logL functions, as well as in the \KpVrest maps. Consequently, we performed several tests changing different parameters controlling the PCA to optimise the algorithm and obtain the best possible telluric removal. In particular, we explore the number of components removed, the spectral slices coadded, and the specific wavelengths (or pixels) used to compute the PCA components.
We note here that our PCA optimisation, differently from other studies in the infrared, is model-independent, i.e. it is not performed by optimising an injected signal but rather by minimising the residual telluric noise. While this is arguably not the best choice to maximise S/N, it is a conservative choice to avoid any optimisation biases as highlighted by e.g. \citet{cabot2019robustness} and \citet{spring2022blackmirror}.
A full comparison with alternative telluric removal methods such as telluric fitting with Molecfit \citep{kausch2015molecfit,smette2015molecfit} or polynomial detrending \citep[e.g.][]{snellen2010HD209458b} is out of the scope of this work and will be the subject of future studies.

% PCA results
Increasing the number of components removed, whether if fixed or variable for each slice, slightly reduces the significance of the telluric residuals, but no improvement is found after removing more than $\sim 6$ components. In all cases, relatively strong telluric residuals remain in the processed data.
As expected, removing the spectral orders that are strongly affected by tellurics from the final coadded \logL results in a reduction of these telluric residuals. However, injection tests show that the injected planetary signal, which is clearly detected using all the orders, even when strong telluric residuals are present, also disappears. This occurs because both telluric and planetary water show the strongest absorption lines (and hence, the strongest signal) in similar spectral ranges.
Finally, we find that modifying the PCA algorithm so that it uses only the specific parts of the spectrum affected by telluric absorption (i.e. pixels that capture telluric lines), rather than using the whole spectral range, results in a significant decrease of the telluric residuals. This happens because, by feeding the algorithm only with telluric-affected regions, telluric-related variations are more noticeable, and hence, are ranked higher than other effects in the PCA components.
Therefore, in our case, avoiding the ranges where tellurics are the strongest in order to mitigate telluric residuals and enhance a potential planetary detection is not a good solution because the planetary signal is also suppressed. Instead, a PCA algorithm fed with pre-defined wavelength ranges where tellurics are known to be present results in a significantly stronger telluric mitigation, whilst preserving any potential \water signals.

% CC grid
We then cross-correlated the spectra resulting from the optimised PCA algorithm with a grid of models covering a range of water abundances and cloud deck levels. We use the CC-to-$\log L$ Bayesian framework which allows us to robustly assess the significance of our results.
We see that models with high water abundances and high cloud deck pressures, and models with low water abundances and low cloud deck pressures are significantly rejected. The preferred models are those with intermediate abundances and cloud deck pressures.
These results are compatible with a potential detection of water in the atmosphere of WASP-166~b. If no water was detected, the preferred models would be those compatible with an almost flat spectrum, i.e., models with low water abundances and low cloud deck pressures, and only models with high water abundance and high cloud deck pressure would be excluded.
We further tried to assess the significance of our best model by computing the CC function with a flat model. Compared to the best model, the flat one is rejected with 4 to 5~$\sigma$, meaning that, regardless of the origin, the data contain a signal $\sim5~\sigma$ above a flat line.

% CC preferred model
In the \KpVrest maps, we observe a correlated signal blue-shifted by about 5~\kms from the expected planetary RV. The signal observed in the two individual nights is shifted from the expected \Kp by a few tens of \kms in opposite directions for each night. However, when both nights are coadded, the signal sits at the expected \Kp and its significance is increased. 
The signals in the individual nights are within 1 and 2~$\sigma$ from the coadded nights signal, meaning that the coadded nights signal is not strongly rejected by the individual night ones. Moreover, the transit observations analysed do not strongly constrain \Kp because they only cover a small part of the total Keplerian orbit.
The shift observed in the planetary \Vrest could be due to winds in the planetary atmosphere. 
Global blue-shifts of the transmission spectrum of hot giant exoplanets have been predicted by several works \citep[e.g.][]{millerriccikempton2012winds,showman2013circulation,rauscher2014circulation}. Such shifts have been observed in the optical through the Na doublet \citep[e.g.][]{wyttenbach2015Na,louden2015Na} and tentatively reported in the infrared through CO \citep{snellen2010HD209458b} and CO and \water \citep{brogi2016rotationwinds,flowers2019GCM}.
The study of the Na doublet at $\sim$589~nm with the same observations as those analysed here \citep{seidel2022wasp166} shows that the Na lines are significantly broadened. This suggests the presence of winds, which seems compatible with what we might be observing here with \water.

% Model process through PCA
An important step in the likelihood-ratio analysis is that the models are processed through the same PCA algorithm as the data. This is necessary to avoid biases introduced by the PCA modifying any potential planetary signal during the telluric correction performed initially, since a PCA can alter both the strength and the shape of the planetary lines, resulting in spurious shifts in \Kp and \Vsys. By processing the models through the same PCA as the data, both data and models should have been modified in the same way, which should result in a better match when performing the cross-correlation. 
In our case, we see that if we use a model without processing it through the same PCA as the data, then the tentative blue-shifted signal is very weak in the second night compared to what we obtain if the model has been adequately processed.
The slow/processed-model method is the only method attempting to reproduce the effects of the telluric removal on the model, and therefore it should be taken as the most reliable reference when quoting a detection. The fast/unprocessed-model method, despite being still common in the literature, is subject to biases with a large variety of telluric-removal algorithms, especially important when retrieving abundances, but also potentially affecting the measured value of \Kp. Therefore, it is not surprising that the two methods give potentially different results, and such discrepancy does not imply that the tentative signal obtained with the slow method cannot be trusted. The biases of the fast/unprocessed-model method have been known for a few years now \citep[e.g. the simulated tests in][]{brogiline2019cc}, and the slow/processed-model approach is standard among many research groups applying Bayesian analysis on high-resolution spectroscopy infrared data \citep[e.g. recently][]{giacobbe2021HD209458giano,line2021wasp77A,gibson2022wasp121,vansluijs2022wasp33}.

% Line lists
To create the grid of models covering several \water abundances and cloud deck pressures we used two different line lists, POKAZATEL and HITEMP, resulting in two sets of 99 models each. Since we are working with ESPRESSO observations, our water models cover optical wavelengths, a range for which published line lists have not been extensively empirically verified, as mentioned above.
% In the visible, water absorption bands are weaker than in the near-infrared. Due to this reduced strength, the accuracy of the model lines in the visible is expected to be worse than in the near-infrared, because their experimental verification is more challenging. 
Hence, we can expect worse accuracy in general and differences between the two different line lists. 
Water lines are known but the line positions are not necessarily accurate. This is key in high-resolution studies such as the one performed here. A lack of accuracy in the line positions could result in Doppler shifts of any expected signal.
Incomplete line lists could make any existing planetary water signal weaker, but we do not expect any possible incompleteness to create shifts in the planetary signal.
In the data analysed here, we see that both the individual \KpVrest maps obtained for each model and the final grid of models are very similar for both POKAZATEL and HITEMP models, which points towards a good agreement between both sets of lines. Despite that, lines could still be inaccurate or incomplete in similar ways, and this agreement does not add evidence to support a planetary origin for the tentative signal observed.

% Shortcomings
We note that when creating the models, we fixed their temperature and scaling factor, and only explored a range of water abundances and cloud deck pressures. We did not consider other sources of opacity. In other words, we did not perform a full atmospheric retrieval and have assumed that the parameters used to create our models are true, because our main goal was only to perform an initial assessment of the presence of water and clouds in the planetary atmosphere. Based on the tentative detection that we obtain, a full atmospheric retrieval is warranted to confirm the results reported here. Further observations of upcoming transits of WASP-166~b could also shed light on the differences obtained between the two nights studied here.

% Conclusion
To summarise, we have analysed the presence of water and clouds in WASP-166~b using two transits observed with ESPRESSO. We use the cross-correlation technique with a grid of models covering a range of water abundances and cloud deck pressures. We find a tentative planetary signal blue-shifted $~$5~\kms from the expected planet velocity in the \KpVrest maps, which could be caused by winds in the atmosphere. A comparison of the different models used favours those with intermediate water abundances and cloud deck pressures. Models with a high water abundance and low cloud deck pressure are strongly rejected, and models with low water abundance and high cloud deck pressure are also not preferred. If no planetary signal was present, we would expect models compatible with a flat spectrum (i.e. low water abundance and high cloud deck pressure) to be favoured, which is not what we observe, hence reinforcing the tentative signal observed in the \KpVrest maps.

%%%%%%%%%%%%%%%%%%%%%%%%%%%%%%%%%%%%%%%%%%%%%%%%%%
%%%%%%%%%%%%%%%%%%%%%%%%%%%%%%%%%%%%%%%%%%%%%%%%%%

\section*{Acknowledgements}

% The Acknowledgements section is not numbered. Here you can thank helpful
% colleagues, acknowledge funding agencies, telescopes and facilities used etc.
This work is based on observations made with ESO Telescopes at the La Silla Paranal Observatory under the programme ID 106.21EM.
ML, HMC, and LD acknowledge funding from a UKRI Future Leader Fellowship, grant number MR/S035214/1. 
SG is grateful to Leiden Observatory at Leiden University for the award of the Oort Fellowship.
RA is a Trottier Postdoctoral Fellow and acknowledges support from the Trottier Family Foundation. This work was supported in part through a grant from FRQNT. 
MLe acknowledges support of the Swiss National Science Foundation under grant number PCEFP2\textunderscore194576. The contribution of MLe has been carried out within the framework of the NCCR PlanetS supported by the Swiss National Science Foundation under grants 51NF40\textunderscore182901 and 51NF40\textunderscore205606. 
We thank G. Frame for useful discussions of this work.
We thank the anonymous referee for the very helpful and thorough review of the article.
% Add acknowledgements
% 
This work made use of \texttt{numpy} \citep{harris2020numpy}, \texttt{scipy} \citep{virtanen2020scipy}, \texttt{astropy} \citep{astropycollaboration2013astropy,astropycollaboration2018astropy}, and \texttt{matplotlib} \citep{hunter2007matplotlib}.

%%%%%%%%%%%%%%%%%%%%%%%%%%%%%%%%%%%%%%%%%%%%%%%%%%
%%%%%%%%%%%%%%%%%%%%%%%%%%%%%%%%%%%%%%%%%%%%%%%%%%
\section*{Data Availability}

% The inclusion of a Data Availability Statement is a requirement for articles published in MNRAS. Data Availability Statements provide a standardised format for readers to understand the availability of data underlying the research results described in the article. The statement may refer to original data generated in the course of the study or to third-party data analysed in the article. The statement should describe and provide means of access, where possible, by linking to the data or providing the required accession numbers for the relevant databases or DOIs.

The ESPRESSO data used in this work is publicly available from the ESO archive under programme ID 106.21EM.

%%%%%%%%%%%%%%%%%%%%%%%%%%%%%%%%%%%%%%%%%%%%%%%%%%
%%%%%%%%%%%%%%%%%%%%%%%%%%%%%%%%%%%%%%%%%%%%%%%%%%



\bibliographystyle{mnras}
\bibliography{wasp166ccf} % if your bibtex file is called example.bib

%\begin{thebibliography}{99}
%\end{thebibliography}

%%%%%%%%%%%%%%%%%%%%%%%%%%%%%%%%%%%%%%%%%%%%%%%%%%

%%%%%%%%%%%%%%%%% APPENDICES %%%%%%%%%%%%%%%%%%%%%

\appendix

\section{PCA implementation details}\label{sec:pcadetails}

Before applying the PCA, the observations of each slice are cleaned from bad pixels. %, centred, and standardised.
% On each observation, we first masked pixels with 0 flux at both ends of the orders.
We corrected for flux anomalies caused by cosmic rays. To do this, we first identified outliers by performing a sigma-clip on values deviating more than $+3$ and $-6$ times the standard deviation of the slice flux (values tailored for these data), and then corrected the identified spikes and the adjacent pixels on each side by linear interpolation of the neighbouring points.
After this, we fitted the continuum of the slice with a linear polynomial and divided by the best fit to remove instrumental slopes.
We also flag channels with low S/N (channels with median flux lower than 2\% of the overall median flux of all channels), which will not be used when computing the PCA eigenvectors.

We then normalised (divided) each observation (each row) by its median flux value. This is done to account for variations in the light throughput in the different observations, so that all of them now have the same baseline flux. 
Each pixel or channel (each column) has its mean subtracted, so that the data matrix is centred. Then each channel is divided by its standard deviation, so that the matrix is now standardised. Hence, each channel has mean equal 0 and a standard deviation of 1.

We note that the previous step of fitting and dividing by a linear polynomial to remove instrumental offsets is not strictly necessary, because any instrumental offset is removed afterwards when standarising each channel. However, removing these instrumental slopes is needed to correctly flag channels with low S/N (if not normalised, the flagged channels would be biased due to the instrumental slope).

% LONG
We then performed a PCA, but instead of directly decomposing the covariance matrix of the data as in \citet{giacobbe2021HD209458giano}, we performed it via a singular value decomposition (SVD) of the standarised data matrix \citep[e.g.][]{dekok2013HD189733}.
If $M$ is the matrix with our standarised data, with dimensions $nf \times nx$ (i.e. rows $\times$ columns), the covariance matrix of the data is then given by $MM^\mathrm{T}/(nf-1)$. This can be diagonalised into $MM^\mathrm{T}/(nf-1)=WDW^\mathrm{T}/(nf-1)$, where $W$ contains the eigenvectors or principal components, and $D$ is a diagonal matrix containing the eigenvalues of a new basis.
We can decompose the data matrix $M$ via an SVD into $M=U \Sigma V^\mathrm{T}$, where $\Sigma$ is an $nf \times nx$ diagonal matrix containing the singular values of $M$, $U$ is an $nf \times nf$ matrix whose columns contain the left singular vectors of $M$, and $V$ is an $nx \times nx$ matrix whose columns contain the right singular vectors of $M$. The singular vectors are a set of orthogonal unit vectors, hence making a new orthonormal basis.
If we now consider $M$ in terms of its SVD, it can be shown that its covariance matrix is $MM^\mathrm{T}/(nf-1) = (U \Sigma V^\mathrm{T}) (U \Sigma V^\mathrm{T})^\mathrm{T}/(nf-1) = (U \Sigma V^\mathrm{T})(V \Sigma U^\mathrm{T})/(nf-1) = U \Sigma^2 U^\mathrm{T}/(nf-1)$.
That is, the singular vectors of $U$ are equivalent to the principal components of the covariance matrix.
% 
% % SHORT
% We then perform a PCA, but instead of directly decomposing the covariance matrix of the data as in \citet{giacobbe2021HD209458giano}, we perform it via a singular value decomposition (SVD) of the standarised data matrix, $M=U \Sigma V^\mathrm{T}$. Here, $\Sigma$ is an $nf \times nx$ (rows $\times$ columns) diagonal matrix containing the singular values of $M$, and $U$ and $V$ are $nf \times nf$ and $nx \times nx$ matrices containing the left and right singular vectors of $M$, respectively. The singular vectors are a set of orthogonal unit vectors, hence making a new orthonormal basis. It can be shown that the singular vectors of $U$ are equivalent to the principal components of the covariance matrix of $M$.

We then created a new matrix containing the first $NC$ columns of $U$ (i.e., the first $NC$ eigenvectors of the SVD of $M$ or, what is the same, the first $NC$ eigenvectors or principal components of the covariance matrix), where $NC$ stands for number of components. %, with an additional column of ones. % $MM^\mathrm{T}$) %We add an additional column of ones.
We used this new matrix to perform a multi-linear regression with the initial matrix of data in order to determine the best-fit coefficients (i.e. the eigenvalues) for the linear combination of the chosen $NC$ components.
This resulted in a fit to the data that should contain most of the telluric, stellar, and instrumental variations, as captured by the first $NC$ components of the PCA.
We then divided the initial matrix of data by this fit and subtract 1. By doing this, we obtained the residuals of the observed data where the telluric, stellar, and instrumental variations captured by the $NC$ components considered have been removed.

Additionally, we applied a high-pass filter in the spectral direction to remove residual instrumental effects from the final processed data. 
Specifically, we first filter out pixels whose scatter (standard deviation) is larger than 2 times the median scatter of that specific pixel in all observations. 
We then fit 2-degree polynomials to each filtered observation to capture any residual instrumental effects, and subtract them out of each observation.
% \textbf{Specifically, we first compute the median scatter (standard deviation) per pixel over all the observations, and all pixels whose scatter is larger than 2 times that median are set to 0.}

%%%%%%%%%%%%%%%%%%%%%%%%%%%%%%%%%%%%%%%%%%%%%%%%%%

\section{Cross-correlation implementation details}\label{sec:ccdetails}

% \subsection{CC approach 1: fast implementation}\label{sec:ccfast}
\subsection{Fast/unprocessed-model CC approach}\label{sec:ccfast}

In this approach, we compute the full CC and \logL functions over a grid of $-100$ to 100~\kms, in steps of 0.5~\kms (which corresponds to the ESPRESSO pixel width). %, i.e., we used a grid of 401 RV shift values. 
That is, we shifted the model to each RV step, interpolated it to the wavelength grid of the observed spectra, and computed the CC and \logL functions following Equations \ref{eq:cc2logL} and \ref{eq:r2cc}.
This is performed slice-by-slice for all the observations of each night, using the different models described in Section~\ref{sec:models}. For a specific model, this results in a CC and \logL function per slice, per observation, and per night.
For each observation, we then combined the \logL functions of each slice by simply coadding them. There is no need to weight the different slices because the \logL already contains information about the different S/N of each slice. This results in a single \logL function per observation, per night.

To enhance the planet signal, the \logL functions of the in-transit observations (where we expect the planetary signal to be) need to be coadded in the planet rest frame. To do this, we shift them by the corresponding planetary orbital velocity $V_\mathrm{p}$, which we computed with the following equation \citep[for which we assume that the planet has no eccentricity,][]{hellier2019wasp166}
\begin{equation}\label{eq:velocity_planet}
% v_\mathrm{p}(t) = V_\mathrm{sys} + K_\mathrm{p} \sin \left( \frac{2 \pi}{P} (t-t_0) \right),
V_\mathrm{p}(t) = V_\mathrm{sys} + K_\mathrm{p} \sin \left[ 2 \pi \varphi (t) \right],
\end{equation}
where \Vsys is the systemic velocity of the system, \Kp is the planet orbital radial velocity semi-amplitude, and $\varphi (t)$ is the planet orbital phase. The phase is defined as
\begin{equation}
\varphi(t) = \frac{t - t_0}{P}
\end{equation}
where $t_0$ is the mid-transit time, and $P$ the orbital period of the planet, so that $\varphi=0$ corresponds to mid-transit.
After shifting all the \logL functions by the corresponding $V_\mathrm{p}$, we only need to coadd them.
When performing the shift to planet rest frame, we spline-interpolated the \logL functions of each observation to a common RV grid. This way, each point of the \logL functions of each observation can be directly summed. This results in a single \logL function per night.

We perform this coadding for different $V_\mathrm{p}$, computed using the expected \Vsys and a range of \Kp from 0~\kms to twice the expected value in steps of 1~\kms (following Equation \ref{eq:velocity_planet}). We obtain the expected \Kp using the following equation with the most up-to-date literature values (see Table \ref{tab:sysparams})
\begin{equation}
% K_\mathrm{p} = \frac{2 \pi a R_\star \sin (i_\mathrm{p})}{P \sqrt{1-e}},
K_\mathrm{p} = \frac{2 \pi}{P} a \sin (i_\mathrm{p}),
\end{equation}
where $a$ is the planet semi-major axis, and $i_\mathrm{p}$, the orbital inclination.
By doing this, we can then produce the usual \KpVsys maps (or \KpVrest if we subtract \Vsys), since the RV grid of the \logL function is equivalent to sampling different \Vsys (see Equation \ref{eq:velocity_planet}). In our case, since we included \Vsys in the computation of $V_\mathrm{p}$, the maps are in the planet rest frame, rather than in the systemic frame.

%%%%%%%%%%%%%%%%%%%%%%%%%%%%%%%%%%%%%%%%%%%%%%%%%%

% \subsection{CC approach 2: precise (and slow) implementation and model processing}\label{sec:ccslow}
\subsection{Slow/processed-model CC approach: precise implementation and model processing}\label{sec:ccslow}

In the second approach, we compute a single CC and \logL value for each pair of \Kp and \Vsys considered. So rather than performing the cross-correlation with the same model shifted by a range of RV steps, we only shift the model once for each pair of \Kp and \Vsys values, and use this shifted model to compute a single point of the CC and \logL functions. %The expected planet RV $V_\mathrm{p}$ is different for each observation and depends on the \Kp and \Vsys values assumed (see Equation \ref{eq:velocity_planet}). Therefore, we have to repeat this operation for each pair of \Kp and \Vsys considered, as we describe in more detail below.
As mentioned before \ref{sec:CCtologL}, this approach allows us to process the model through the same PCA as the data prior to performing the cross-correlation. This is key to avoid biases and should result in a better match between model and data.
%
% The main advantage of this approach is that it allows us to process the model used in the CC through the same PCA as the data, which is not possible in the first approach.
% This is important because the PCA might alter the planetary signal contained in the data. The models we used in the first approach do not contain any change due to the PCA, therefore, the match with a possible planetary signal will not be as good as if the model has also been altered in the same way as the data. Due to this mismatch, our planetary detection might be weaker. Moreover, when performing the model comparison (see below Section \ref{sec:modelcomparison}), we could also misinterpret the water abundance and cloud deck pressure because the line depths do not match between model and data.
% Hence, we also processed the model through the same PCA applied to the data. This is done before computing the CC and \logL values.
In the following, we explain this model processing and the computation of the \logL with this slow/processed-model approach.

To process the planetary water model through the same PCA as the data, we first created a data matrix with the same dimensions as the original spectra ($nf \times nx$) containing the model that will be used to compute the CC (instead of the observed data).
% Again, we are working order-by-order, so we repeated this process for each order.
For the rows corresponding to in-transit observations, the model matrix contains the model shifted to the expected planet RV, interpolated to the same wavelength grid as each observation. For the rows corresponding to out-of-transit observations, the data matrix contains only ones.

As explained in Section \ref{sec:pca}, the linear regression of the data with the selected PCA components results in a fit matrix that should only contain (if the PCA works as expected) the fitted tellurics and stellar lines, and changes in flux due to varying airmass and throughput. It also contains the overall drop in flux due to the planet transit, i.e., the broadband transmission planet spectrum. We want to inject the planetary water model to this fit matrix so that the model contains the same variability as the data. However, our model matrix already contains the drop in flux due to the planet transit, because the models are expressed in units of $1-(R_\mathrm{p}/R_\star)^2$. Therefore, before injecting the models, we need to normalise them to remove this effect. We do this by dividing the in-transit observations in the model matrix by their mean (we do not need to apply any change to the out-of-transit observations, which are simple a flat spectrum at flux one). After this, we injected the normalised model to the fit from the data by multiplying the two matrices.

We then apply the full PCA processing to this last matrix (including the out-of-transit observations), as done originally with the data. That is, we use the same number of components and bad-pixel masks, and perform the centering and standarisation, singular value decomposition, and linear regression. This results in a matrix with the processed model per observation, which should have been altered by the PCA in the same way as the real data.

We then computed the CC and \logL of the in-transit observations using the same method as in the first approach (i.e. Equations \ref{eq:cc2logL} and \ref{eq:r2cc}). In this case, however, we have already shifted the template to the expected planet RV (for a specific pair of \Kp and \Vsys values). Therefore, we only compute the CC and \logL once for each observation. To get a single \logL value per observation, we then directly sum the \logL values that we obtain for each slice.
Since the \logL has been computed with the model already shifted to the expected planet RV, we can directly sum the \logL of all the observations, as we are already in planet rest frame, and there is no need for interpolation as in the first approach. This directly gives us a data point on the \KpVsys maps. The processing of the model depends on the chosen \Kp and \Vsys values, therefore, we repeated this whole process (model processing and computation of a single CC and \logL value) for each pair of \Kp and \Vsys values considered, resulting in the full \KpVsys map (or again, \KpVrest if we subtract \Vsys).

%%%%%%%%%%%%%%%%%%%%%%%%%%%%%%%%%%%%%%%%%%%%%%%%%%

\section{POKAZATEL models for high water abundances}\label{sec:models_mu}

\begin{figure*}
\centering
\includegraphics[width=\textwidth]{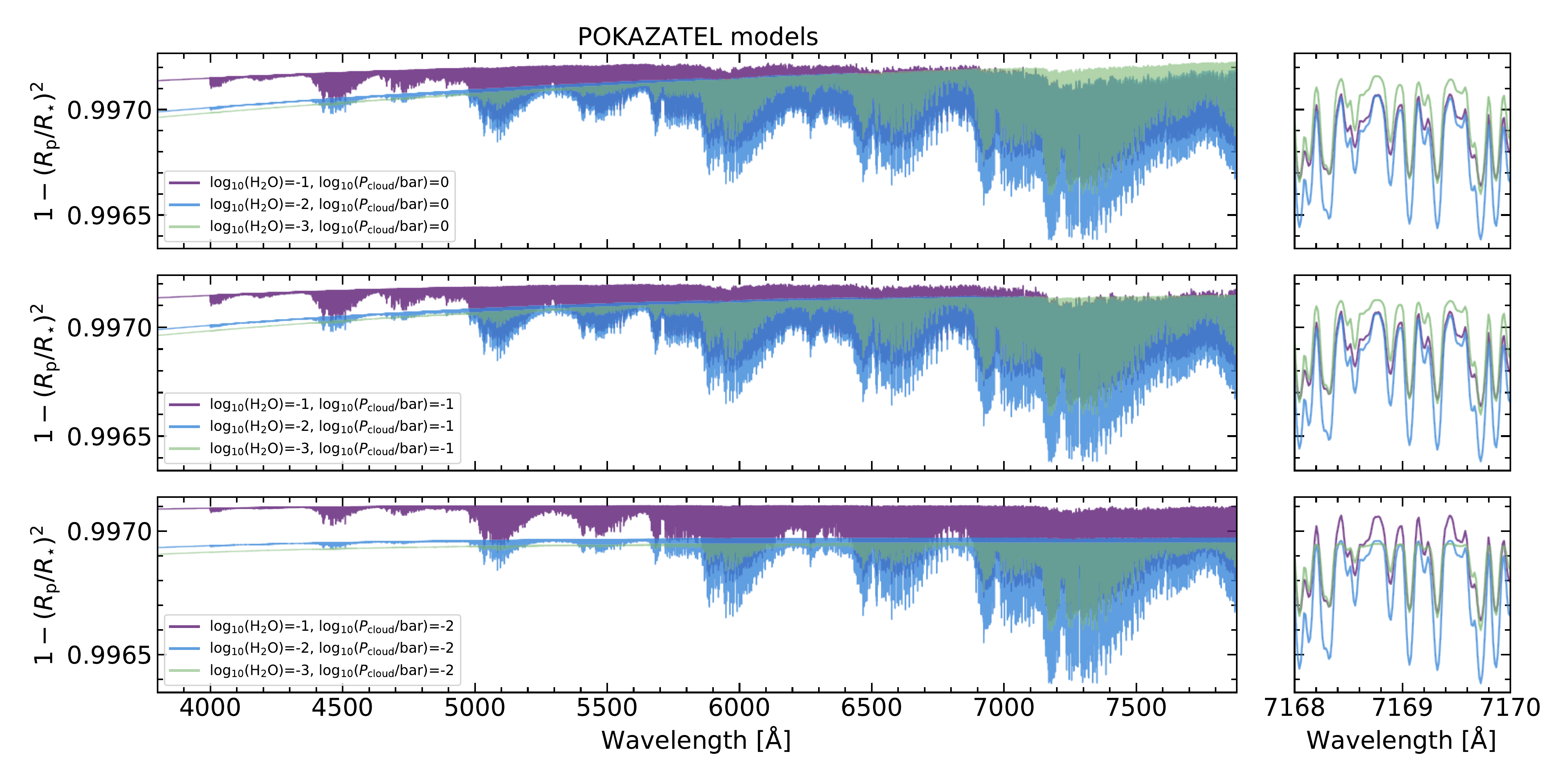}
\caption{Left: POKAZATEL \water templates for WASP-166~b covering the ESPRESSO wavelength range for a range of cloud deck pressures (\logcloud$=0$ top, \logcloud$=-1$ middle, \logcloud$=-2$ bottom), and a range of water abundances (depicted by the various colours in all panels). Right: Zoom in on a region with strong absorption lines. This figure shows the decrease in absorption strength for water rich atmospheres (\logwater$=-1$, purple) compared to lower abundances (\logwater$=-2$, blue) due to the increase in mean molecular weight for water-rich atmospheres (see Section \ref{sec:results_gridsearch}). For lower water abundances (\logwater$=-3$, green), the strength of the absorption features decreases as expected due to the decrease in water content.}
\label{fig:models_mu}
\end{figure*}

%%%%%%%%%%%%%%%%%%%%%%%%%%%%%%%%%%%%%%%%%%%%%%%%%%

% Don't change these lines
\bsp	% typesetting comment
\label{lastpage}
\end{document}